\documentclass[a4paper,11pt]{article}
\pdfoutput=1 

\usepackage{jcappub} 
\usepackage{amsmath}
\usepackage{amssymb}
\usepackage[T1]{fontenc} 

\title{\boldmath  \textit{Interacting dark energy: clarifying the cosmological implications and viability conditions}}

\author[a]{Marcel A. van der Westhuizen}
\author[a, b]{and Amare Abebe}

\affiliation[a]{Centre for Space Research, North-West University,\\ Potchefstroom 2520, South Africa}
\affiliation[b]{National Institute for Theoretical and Computational Sciences (NITheCS),\\ South Africa}
\emailAdd{marcelvdw007@gmail.com}
\emailAdd{Amare.Abebe@nithecs.ac.za}

\abstract{In this study, cosmological models are considered, where dark matter and dark energy are coupled and may exchange energy through non-gravitational interactions with one other. These interacting dark energy (IDE) models have previously been introduced to address problems with the standard $\Lambda$CDM model of cosmology (which include the coincidence problem, Hubble tension and $S_8$ discrepancy). However, conditions ensuring positive energy densities have often been overlooked. Assuming two different linear dark energy couplings, $Q = \delta H \rho_{\rm{de}}$ and $Q = \delta H \rho_{\rm{dm}}$, we find that negative energy densities are inevitable if energy flows from dark matter to dark energy (iDMDE regime) and that consequently, we should only seriously consider models where energy flows from dark energy to dark matter (iDEDM regime). To additionally ensure that these models are free from early time instabilities, we need to require that dark energy is in the `phantom' ($\omega<-1$) regime. This has the consequence that model $Q=\delta H \rho_{\rm{dm}}$ will end with a future big rip singularity, while $Q = \delta H \rho_{\rm{de}}$ may avoid this fate with the right choice of cosmological parameters.}

\keywords{dark energy theory, dark matter theory, cosmology of theories beyond the SM}

\makeatletter
\gdef\@fpheader{}
\makeatother

\begin{document}
\maketitle
\flushbottom
\newpage
\section{Introduction}\label{sec:intro}
The expansion of the universe has thus far been well described by the $\Lambda$CDM model, where the energy budget of the universe is divided between $\approx 5 \%$ baryonic matter (standard model particles), $\approx 25 \%$ non-baryonic cold dark matter (which keeps galaxies from flying apart) and $\approx 70 \%$ dark energy in the form of the cosmological constant $\Lambda$ (which explains late-time accelerated expansion). This model has proven to be a very successful fit for astrophysical observations \cite{Planck2018, Riess1998, Perlmutter1999, kids2021, DES2022}, but problems with the $\Lambda$CDM model remain, which include:
\begin{itemize}
 \item \textit{The Cosmological Constant Problem} or vacuum catastrophe, refers to the measured energy density of the vacuum being over $120$ orders of magnitude smaller than the theoretical prediction. This has been referred to as the worst prediction in the history of physics and casts doubt on dark energy being a cosmological constant \cite{Weinberg1989, HobsonTextbook}. 
\item \textit{The Cosmic Coincidence Problem}, which alludes to the dark matter and dark energy densities having the same order of magnitude at the present moment while differing with many orders of magnitude in the past and predicted future \cite{delCampo2009, Huey2006, Velten2014, Wang2016, Bolotin2015}. The initial conditions of dark matter and dark energy should be fine-tuned to about $95$ orders of magnitude to produce a universe where the two densities nearly coincide today, approximately $14$ billion years later \cite{Zlatev1999}.
\item \textit{The Hubble Tension}, which concerns the $\sim 5 \sigma$ level difference between values of the Hubble constant $H_0$ as measured from the Cosmic Microwave Background (CMB) versus the value obtained from Type Ia Supernovae using a calibrated local distance ladder \cite{Planck2018, Riess2019, Riess2021, ValentinoH02017, ValentinoH02020, ValentinoH02021, Cyr-Racine2021, Nunes2022, Gariazzo2022,VagnozziV22020,VagnozziV22018, LWang02022, Dainotti2021, Dainotti2022}.
\item \textit{The $S_8$ discrepancy}, which concerns the $3 \sigma$ level difference between measurements made from the CMB against weak lensing measurements and redshift surveys of the parameter $S_8$, which quantifies the amplitude of late-time matter fluctuations and structure growth. \cite{LWang02022, Lucca2021, NunesS82021, ValentinoS82021, Gariazzo2022, VagnozziV22018}.
\end{itemize}
These problems motivate research into new physics beyond the $\Lambda$CDM model. A popular approach to solving these problems has been to investigate cosmological models in which there are non-gravitational interactions between the dark sectors of the universe. This allows the two dark sectors to exchange energy (and/or momentum) while dark matter (DM) and dark energy (DE) are not separately conserved, but the energy (and momentum) of the total dark sector is conserved instead. These models are broadly known as interacting dark energy (IDE) models. In these models, we assume that both radiation and baryonic matter is uncoupled and separately conserved since there are strong 'fifth-force' observational constraints on baryonic matter \citep{Carroll2021} and any new significant interactions with photons would probably cause deviations from photons following a geodesic path \citep{Wang2016}. We, therefore, have the following conservation equations for interacting dark energy models:
\begin{gather} \label{IDE1.3}
\begin{split}
\dot{\rho}_{\rm{dm}} + 3 H\rho_{\rm{dm}} = Q \quad \quad &; \quad \quad \dot{\rho}_{\rm{de}} + 3 H\rho_{\rm{de}} (1 + \omega) = -Q\; , \\
\dot{\rho}_{\rm{bm}} + 3 H\rho_{\rm{bm}} = 0 \quad \quad &; \quad \quad \dot{\rho}_{\rm{r}} + 3 H\rho_{\rm{r}} (1 + 1/3) = 0\;,
\end{split}
\end{gather}
where $H$ is the Hubble parameter, $\rho$ is the energy density, and the subscripts denote radiation (r), baryonic matter (bm), dark matter (dm) and dark energy (de). Here we still assume pressure-less dark matter ($P_{\rm{dm}}=0 \rightarrow \omega_{\rm{dm}}=0$), and note that baryonic matter $\omega_{\rm{bm}}=0$ and radiation $\omega_{\rm{r}}=1/3$ are uncoupled $Q=0$, as in the $\Lambda$CDM model. The dark energy equation of state ($\omega_{\rm{de}}=\omega$ from here onwards) is left as a free variable since the dark energy may be either vacuum energy ($\omega=-1$), in the quintessence ($-1<\omega<-1/3$) or phantom ($\omega<-1$) regime. Here $Q$ is an arbitrary coupling function whose sign determines how energy (or momentum) is transferred between dark energy and dark matter. If $Q>0$, then the energy (or momentum) is transferred from dark energy to dark matter and vice versa for $Q<0$, such that  \citep{Wang2016, Bolotin2015, ValentinoH02020, Vliviita2009, Bhmer2008, Gavela2009, Gavela2010, ValentinoCU2021, ValentinoCT2020, Pan2020, Lucca2020, Lucca2021}:
\begin{equation} \label{IDE1.4}
Q =
  \begin{cases}
    >0 & \text{Dark Energy $\rightarrow$ Dark Matter (iDEDM regime)}\\
    <0 & \text{Dark Matter $\rightarrow$ Dark Energy (iDMDE regime)}\\    
    =0 & \text{No interaction}.  \vspace{-0.0cm}
  \end{cases}     
\end{equation}
Here we have denoted the interacting case where energy flows from dark energy to dark matter ($Q>0$) as the interacting Dark Energy Dark Matter regime (iDEDM), and vice versa as the interacting Dark Matter Dark Energy regime (iDMDE) \citep{Lucca2021}. Since there is currently no fundamental theory for the coupling equation $Q$, the coupling in most works is purely phenomenologically motivated; and must be tested against observations \citep{Wang2016, Bolotin2015, ValentinoH02020, Vliviita2009, Bhmer2008, Gavela2009, Gavela2010, ValentinoCU2021,ValentinoCT2020, Pan2020, Lucca2020, Lucca2021}.
The coupling is thus freely chosen, but we will only consider models where the coupling function $Q$ is either proportional to the dark matter or the dark energy density, which could have a strong field theoretical ground \citep{PanField2020}. The core publications we considered for these models are \citep{Vliviita2009, Gavela2009, Gavela2010}. For recent developments and observational constraints, see \citep{ValentinoH02017, ValentinoH02020, ValentinoCU2021, ValentinoCT2020, Pan2020, Lucca2020, Lucca2021, PanField2020, LWang02022, NunesS82021, Gariazzo2022}; and for comprehensive review articles on interacting dark energy, see \citep{Wang2016, Bolotin2015}. \\
IDE models were first introduced to address the coincidence problem \cite{delCampo2009, Huey2006, Velten2014, Wang2016, Bolotin2015}, but this approach has recently become less popular due to observational constraints on the interaction strength needed to solve this problem significantly. Instead, in recent years these models have received more attention as possible candidates to alleviate the Hubble tension \cite{ValentinoH02017, ValentinoH02020, ValentinoH02021, Cyr-Racine2021, LWang02022, Nunes2022, Gariazzo2022,VagnozziV22018} and have most recently been shown also to alleviate the $S_8$ discrepancy while making an excellent fit to the latest cosmological data available \cite{Lucca2021, LWang02022, VagnozziV22018, NunesS82021, Gariazzo2022}. \\
Even though IDE models have proven to be popular candidates to address the biggest problems in cosmology, we believe that the parameter space in the most popular model is often not well understood as conditions to ensure positive energy densities and to avoid a future big rip are often ignored. We would like to clarify this. \textit{New results that we obtained will be shown in italics to clearly differentiate from previously known results}.  
In this article:
\begin{itemize}
\item We clarify the general cosmological implications of IDE models for any interaction $Q$ (summarised in table \ref{Table_Q_events_relative}).
\item We derive \textit{a new equation (\ref{DSA.8}) that may be used to easily obtain phase portraits for the evolution of dark matter and dark energy densities (figure \ref{PP_Q2}), which applies to any interaction $Q$ without the need to solve the conservation equation (\ref{IDE1.3})}, which may be excessively difficult for most interaction functions.
\item We analyze the most popular IDE model where the interaction is proportional to the dark energy density $Q = \delta H \rho_{\rm{de}}$. For this model, \emph{we will derive the viability condition $ 0 <   \delta  < - 3 \omega/( 1 + \frac{1}{r_0})$ (\ref{Q2_PEC.5}) to avoid negative energy densities, which is the most important result of this paper, as this is nearly always ignored in the literature and shows that negative energy densities are inevitable if energy flows from dark matter to dark energy (iDMDE regime) and that consequently, we should only seriously consider models where energy flows from dark energy to dark matter (iDEDM regime)}. Most of the rest of the paper is dedicated to clarifying previous results of this model while showing how this negative energy problem is always present for the iDMDE regime's future expansion.
\item For this model, we show that the iDEDM regime may alleviate the coincidence problem for the past expansion while solving the coincidence problem for the future expansion, as seen in \ref{eos.summary_Q2} and figure \ref{Coincidence_Problem_Q2}. Conversely, the iDMDE regime is shown only to worsen the coincidence problem. \textit{This result is well known, but as an application of the previous point, we included newly extended plots for the future expansion which clearly show negative values of $r$, indicating negative energy densities.}
The way we have related the effective equations of state $\omega^{\rm{eff}}_{\rm{dm}}$ and $\omega^{\rm{eff}}_{\rm{de}}$ to various parameters in this section and throughout the rest of the paper are not new results, \textit{but rather a newly suggested formulation for analyzing familiar problems and results more clearly.}
\item We clarify the additional cosmological consequences of the iDEDM and iDMDE regime for this model, using imposed parameters for the sake of clarity. We show how the coupling influences the evolution of the energy densities (figure \ref{fig_rho_Q2} and \ref{fig_Omega_Q2}), the total fluid effective equation of state (figure \ref{omega_eff_evo_Q2}), deceleration parameter (figure \ref{q_evo_Q2}), the expansion rate (figure \ref{H_rel_Q2}), and the age of the universe (figure \ref{scalefactor_Q2}). Most of these results have been shown before, \textit{but we included newly extended plots for the future expansion clearly showing how negative dark matter energy densities plague all these previously known results for the iDMDE regime.}
\item \textit{We derived new analytical expressions for the redshift where the radiation-matter $z_{(\rm{r=dm+bm})}$ (\ref{EQ_Q2_3}) and matter-dark energy $z_{(\rm{dm+bm=de})}$  (\ref{EQ_Q2_6}) equalities occur as well the transition redshift $z_{\rm{t}}$ (\ref{EQ_Q2_10}) where accelerated expansion starts, known as the cosmic jerk.} These results, with the imposed parameters, are summarised in tables \ref{Table_LCDM_events3}, \ref{Table_Q2_iDEDM_events} and \ref{Table_Q2_iDMDE_events}.
\item We show that to ensure that this model is not only free from negative energy densities but also free from early time instabilities, \textit{we need to require that dark energy is in the `phantom' ($\omega<-1$) regime, as shown in table \ref{Tab_PECStabQ2}. This has the consequence that these universe models will end with a future big rip singularity (figure \ref{a_evo_rip_Q2}) unless the effective equation of state is $\omega^{\rm{eff}}_{\rm{de}} = \omega+\frac{\delta}{3}>-1$. The results from table \ref{Tab_PECStabQ2} and equation (\ref{AllCON}) give the most useful theoretical constraints on parameters for other researchers who want to constrain these models with new observational data. A new equation (\ref{BR.5_Q2}), which gives the exact time of this big rip, has also been derived}.
\item We briefly summarise the same results above for the interaction function $Q = \delta H \rho_{\rm{dm}}$. Here the positive energy density condition $0 <   \delta  < -  \frac{3 \omega}{\left( 1 + r_0 \right) }$ (\ref{Q1_PEC.5}) and table \ref{Tab_PECStabQ1} summarises the most useful theoretical constraints for this model. The negative energy densities and the inevitable big rip for this model are well known, thus \textit{all results in this section may be considered as only new formulations of known results}. \textit{Equation (\ref{BR.5_Q1}) giving the time of this big rip is the only new result}.  
\end{itemize}
\newpage

\section{Properties of interacting dark energy models}
\subsection{Background cosmology and its implications}\label{sec: distance modulus}
In IDE cosmology, the standard assumptions of isotropy and homogeneity of the universe, as characterised by the Friedmann-Lema\^{i}tre-Robertson-Walker (FLRW) metric, still hold. At the same time, only the conservation equations are modified, changing the evolution of the DM and DE density profiles. This implies that the standard $\Lambda$CDM equations may be used, which include those for the Friedmann equation, the deceleration parameter and the total effective equation of state, respectively:
\begin{align}
 H^2(a) &= \left( \frac{\dot{a}}{a} \right)^2  = \frac{8 \pi G}{3} \left(\rho_{\rm{r}} + \rho_{\rm{bm}}+ \rho_{\rm{dm}} +\rho_{\rm{de}} \right)  -\frac{kc^2}{a^2},  \label{H} \\
q &=    \Omega_{\rm{r}}  + \frac{1}{2} \left( \Omega_{\rm{bm}}+ \Omega_{\rm{dm}} \right)  + \frac{1}{2} \Omega_{\rm{de}}  \left(1 +3 \omega \right), \label{q}\\
\omega^{\rm{eff}} &= \frac{P_{\rm{tot}}}{\rho_{\rm{tot}}} = \frac{\frac{1}{3}\Omega_{\rm{r}} + \omega_{\rm{de}}\Omega_{\rm{de}}}{\Omega_{\rm{r}}+\Omega_{\rm{bm}}+\Omega_{\rm{dm}}+\Omega_{\rm{de}}}\;, \label{eff}
\end{align}
where $\omega_{\rm{dm}}=\omega_{\rm{bm}}=0$ and $\omega_{\rm{r}}=1/3$. The crucial difference in the behaviour of IDE models may be understood by how the interaction affects the effective equations of state of both dark matter $\omega^{\text{eff}}_{\text{dm}}$ and dark energy $\omega^{\text{eff}}_{\text{de}}$, relative to the uncoupled background equations ($Q=0$) in {(\ref{IDE1.3})}  such that \citep{Vliviita2009, Gavela2009, Bolotin2015}:
\begin{gather} \label{3}
\begin{split}
\omega^{\text{eff}}_{\text{dm}} = - \frac{Q}{3 H \rho_{\text{dm}}} \qquad ; \qquad  \omega^{\text{eff}}_{\text{de}} =  \omega_{\text{de}} + \frac{Q}{3 H \rho_{\text{de}}} \;. \\
\end{split}
\end{gather}
Thus, the effects of an interaction may be understood to imply that if \citep{Vliviita2009, Gavela2009, Bolotin2015}: \\
\begin{equation} \label{IDE1.10}
Q > 0 \text{ (iDEDM)}
  \begin{cases}
   \omega^{\rm{eff}}_{\rm{dm}} <0 & \text{Dark matter redshifts \textit{slower} than a}^{-3}  \text{ (less DM in past)}, \\
    \omega^{\rm{eff}}_{\rm{de}}> \omega_{\rm{de}} & \text{Dark energy has \textit{less} accelerating pressure}, \\   
  \end{cases}     
\end{equation} \vspace{-0.5cm}
\begin{equation} \label{IDE1.11}
Q  < 0 \text{ (iDMDE)}
  \begin{cases}
   \omega^{\rm{eff}}_{\rm{dm}} >0 & \text{Dark matter redshifts \textit{faster} than a}^{-3}  \text{ (more DM in past)}, \\
    \omega^{\rm{eff}}_{\rm{de}}< \omega_{\rm{de}} & \text{Dark energy has \textit{more} accelerating pressure}. \\   
  \end{cases}     
\end{equation}
This implies that even if $\omega_{\rm{de}}=-1$, when $Q<0$ or $Q>0$, then the dark energy may behave like either uncoupled quintessence $\omega^{\rm{eff}}_{\rm{de}}>-1$  or uncoupled phantom $\omega^{\rm{eff}}_{\rm{de}}< -1$ dark energy respectively. If there is no interaction between dark matter and dark energy $(Q=0)$, the effective equations of state reduce back to the uncoupled model, such that $\omega^{\rm{eff}}_{\rm{dm}} = \omega_{\rm{dm}} = 0$ and $\omega^{\rm{eff}}_{\rm{de}}  = \omega_{\rm{de}}$. \\
These effective equations of state allow us to make predictions regarding the consequences of a dark coupling; and why it was initially introduced to address the cosmic coincidence problem. This can be seen by considering the ratio $r_{\rm{IDE}}$ of $\rho_{\rm{dm}}$ to $\rho_{\rm{de}}$ from (\ref{IDE1.3}) for interacting dark energy models:
\begin{gather} \label{IDE1.12}
\begin{split}
r_{\rm{IDE}} &= \frac{\rho_{\rm{dm}}}{\rho_{\rm{de}}} = \frac{  \rho_{\rm{(dm,0)}} a^{-3(1+\omega^{\rm{eff}}_{\rm{dm}}})  }{ \rho_{\rm{(de,0)}} a^{-3(1+\omega^{\rm{eff}}_{\rm{de}}})}  = r_0 a^{-\zeta_{\rm{IDE}}} \quad \quad ; \quad \quad \text{with}  \quad \zeta_{\rm{IDE}}= 3 \left(\omega^{\rm{eff}}_{\rm{dm}} - \omega^{\rm{eff}}_{\rm{de}} \right),  \\ 
\end{split}
\end{gather}
with $\zeta$ indicating the magnitude of the coincidence problem. Thus, from (\ref{IDE1.12}), we see that the smaller the difference between $ \omega^{\rm{eff}}_{\rm{dm}}$ and $\omega^{\rm{eff}}_{\rm{de}}$ the more the coincidence problem will be alleviated while being solved if $\zeta=0$, which happens when $\omega^{\rm{eff}}_{\rm{dm}} = \omega^{\rm{eff}}_{\rm{de}}$. This can be achieved if dark matter redshifts slower $\omega^{\rm{eff}}_{\rm{dm}} <\omega_{\rm{dm}}$ and dark energy redshifts faster $\omega^{\rm{eff}}_{\rm{de}}>\omega_{\rm{de}}$, which coincides with the iDEDM ($Q>0$) scenario.
The opposite holds for the iDMDE ($Q<0$) scenario.  From (\ref{IDE1.10}),(\ref{IDE1.11}) and (\ref{IDE1.12}), while noting that $\zeta_{\Lambda\rm{CDM}}=3$, we may conclude:
\begin{equation} \label{CCP.5}
\zeta_{\rm{IDE}}= 3 \left(\omega^{\rm{eff}}_{\rm{dm}} -  \omega^{\rm{eff}}_{\rm{de}} \right)
 \begin{cases}
   Q>0\text{ (iDEDM):} \quad \zeta_{\rm{IDE}} < \zeta_{\Lambda\rm{CDM}} & \text{\textit{alleviates} coincidence problem}, \\
   Q<0\text{ (iDMDE):} \quad \zeta_{\rm{IDE}} > \zeta_{\Lambda\rm{CDM}} & \text{\textit{worsens} coincidence problem}. \\
 \end{cases}    
\end{equation}
Besides addressing the coincidence problem, IDE models have other far-reaching cosmological consequences. Since $\omega^{\rm{eff}}_{\rm{dm}}<0$ for iDEDM, DM redshifts \textit{slower}, which leads to \textit{less} DM in the past and the radiation-matter equality happening \textit{later} \citep{Gavela2009}, which in turns causes suppression in the matter power spectrum, \textit{alleviating} the $S_8$ discrepancy \citep{Lucca2021} (see \citep{VagnozziV22018, NunesS82021, Gariazzo2022,Lucca2021, LWang02022} for how IDE models address the $S_8$ tension). Similarly, $\omega^{\rm{eff}}_{\rm{de}}>\omega_{\rm{de}}$, such that DE redshifts faster, causing \textit{more} DE in the past. Less DM and more DE in the past have the consequence that both the cosmic jerk and the matter-dark energy equality happen \textit{earlier} in cosmic history. From the Friedmann equation (\ref{H}), we can see that this overall suppression of dark matter density causes a lower value of the Hubble parameter at late times. This lower value of $H_0$ worsens the Hubble tension with regard to late time probes \citep{Lucca2021} (see \citep{ValentinoH02020, ValentinoH02021, Cyr-Racine2021, Nunes2022, Gariazzo2022, VagnozziV22018, LWang02022} for how IDE models address the Hubble tension). Since the Hubble parameter, and therefore the expansion rate, is lower throughout most of the expansion, the universe must have taken longer to reach its current size. Since more time was needed to reach current conditions, the universe should also be \textit{older}. The opposite holds for the iDMDE scenario. These general consequences of a dark sector coupling (if all other parameters are kept constant), are summarised in table \ref{Table_Q_events_relative}.
\begin{table}[h]
\caption{\label{Table_Q_events_relative} Consequences of interacting dark energy models (relative to uncoupled models)} 
\begin{center}
\begin{tabular}{|c|c|c|}
\hline
\rule{0pt}{13pt}  &  $Q>0$  &  $Q<0$ \\
\hline
\hline
\rule{0pt}{13pt} Energy flow & DE $\rightarrow$ DM \;(iDEDM) & DM $\rightarrow$ DE  \;(iDMDE) \\
\hline
\rule{0pt}{13pt} Effective equations of state & $\omega^{\rm{eff}}_{\rm{dm}}< \omega_{\rm{dm}}$\; ; \;$\omega^{\rm{eff}}_{\rm{de}}> \omega_{\rm{de}}$ & $\omega^{\rm{eff}}_{\rm{dm}}> \omega_{\rm{dm}}$ \;;\; $\omega^{\rm{eff}}_{\rm{de}}< \omega_{\rm{de}}$ \\
\hline
\rule{0pt}{13pt} Coincidence problem & Alleviates ($\zeta_{\rm{IDE}} < \zeta_{\Lambda\rm{CDM}}$) & Worsens ($\zeta_{\rm{IDE}} > \zeta_{\Lambda\rm{CDM}}$) \\
\hline
\rule{0pt}{13pt} Hubble tension & Worsens  & Alleviates \\
\hline
\rule{0pt}{13pt} $S_8$ discrepancy & Alleviates  & Worsens \\
\hline
\rule{0pt}{13pt} Age of universe & Older & Younger   \\
\hline
\rule{0pt}{13pt} Radiation-matter equality & Later ($z_{\rm{IDE}}<z_{\Lambda\rm{CDM}}$)  &  Earlier ($z_{\rm{IDE}}>z_{\Lambda\rm{CDM}}$) \\
\hline
\rule{0pt}{13pt} Cosmic jerk ($q=0$) & Earlier ($z_{\rm{IDE}}>z_{\Lambda\rm{CDM}}$) & Later ($z_{\rm{IDE}}<z_{\Lambda\rm{CDM}}$)\\
\hline
\rule{0pt}{13pt} Matter-dark energy equality & Earlier ($z_{\rm{IDE}}>z_{\Lambda\rm{CDM}}$) & Later ($z_{\rm{IDE}}<z_{\Lambda\rm{CDM}}$) \\
\hline
\end{tabular}
\end{center}
\end{table} \\  
These implications will only hold if the IDE model is viable. Any cosmological model may be considered unviable due to theoretical concerns, such as internal inconsistencies, instabilities or negative energy densities. A model free of these problems can only be deemed viable if it meets observational constraints, such as predicting an expansion history that coincides with the most recent cosmological data. This paper will only consider theoretical constraints, while we refer readers to  \citep{ValentinoH02017, ValentinoH02020, ValentinoH02021, ValentinoCU2021, ValentinoCT2020, Pan2020, Lucca2020, Lucca2021, PanField2020, NunesS82021, Gariazzo2022, VagnozziV22018, Cyr-Racine2021, LWang02022, Nunes2022} for observational constraints.  
\subsection{Instabilities and the doom factor} \label{doom}
The coupling between the dark sectors will influence the evolution of dark matter and dark energy perturbations. A complete perturbation analysis of the models considered in the paper is found in \citep{Vliviita2009} and \citep{Gavela2009}. For our purposes, we only want to know what combination of parameters may be used to avoid instabilities. This can be found in \citep{Gavela2009}, by introducing the so-called doom factor \textbf{d}:
\begin{gather} \label{d.1}
\begin{split}
\textbf{d} = \frac{Q}{3 H \rho_{\rm{de}} (1+\omega)}.
\end{split}  	
\end{gather}
This is called the doom factor since this factor is proportional to the coupling function $Q$ and may induce non-adiabatic instabilities in the evolution of the dark energy perturbations \citep{Gavela2009}. The sign of \textbf{d} will determine if there is an early time instability. It was shown that if the doom factor is positive and large $\textbf{d}>1$; the dark energy perturbations will become dominated by the terms which are dependent on the coupling function $Q$, leading to a runaway; unstable growth regime \citep{Gavela2009}. As long as $\textbf{d}<0$, the model should be free of non-adiabatic instabilities at large scales. This doom factor can therefore provide the range of parameters that will give a priori stable universe, as is often done in literature \citep{Gavela2009, ValentinoH02020, ValentinoH02021, Cyr-Racine2021, Gariazzo2022, Lucca2020, Lucca2021,Nunes2022, VagnozziV22018}. \\ 
\subsection{Evolution of energy densities and phase portraits} \label{Evolution of energy densities and phase portraits}
Since the coupling function, $Q$, is phenomenologically motivated, many different interaction functions exist, which could either be simple linear or complex non-linear interactions \citep{Wang2016, Bolotin2015}. This often leads to difficulties when trying to solve the coupled conservation equations (\ref{IDE1.3}) to obtain analytical expressions for how $\rho_{\rm{dm}}$ and $\rho_{\rm{de}}$ evolve. It is, therefore, informative to consider how the derivatives of the density parameters $\dot{\Omega}_{\rm{x}}$ evolve for any arbitrary coupling $Q$. This can be used to obtain phase portraits with flow lines in the ($\Omega_{\rm{dm}}$, $\Omega_{\rm{de}}$)-plane that has attractor and repulsor points. These attractor and repulsor points can tell us how the ratio of DM to DE evolves and indicate whether the coupling solves the coincidence problem. Furthermore, these phase portraits can also tell us if the DM or DE energy densities become negative at any point, which indicates that the interaction $Q$ is unphysical. \\ 
For this analysis, we will consider models which contain radiation $\Omega_{\rm{r}}$, baryonic matter $\Omega_{\rm{bm}}$, dark matter $\Omega_{\rm{dm}}$ and dark energy $\Omega_{\rm{de}}$. This may be done by first considering how the density parameters evolve with time, which is done by taking the derivative of $\Omega_{\rm{x}}=\frac{8 \pi G }{3 H^2} \rho_{\rm{x}}$, (where x can be either r, bm, dm or de), giving:
\begin{gather} \label{DSA.1}
\begin{split}
\dot{\Omega}_{\rm{x}} = \dot{\left[\frac{8 \pi G }{3 H^2} \rho_{\rm{x}} \right]}   &= \frac{8 \pi G }{3} \left[ \frac{\dot{\rho_{\rm{x}}}}{H^2} - \rho \frac{2\dot{H}}{H^3} \right] = \frac{8 \pi G }{3H^2} \left[\dot{\rho_{\rm{x}}} - \rho \frac{2\dot{H}}{H} \right]. \\
\end{split}
\end{gather}
From the conservation equations for interacting dark energy models (\ref{IDE1.3}), we have:
\begin{gather} \label{DSA.2}
\begin{split}
\dot{\rho}_{\rm{x}} &= - 3 H \rho_{\rm{x}} (1 + \omega_{\rm{x}}) \pm Q, \\
\end{split}
\end{gather}
where $\pm=+$ for $x=$dm and $\pm=-$ for $x=$de. Substituting (\ref{DSA.2}) into (\ref{DSA.1}) gives:
\begin{gather} \label{DSA.3}
\begin{split}
\dot{\Omega}_x &= \frac{8 \pi G }{3H^2} \left[ - 3 H \rho_{\rm{x}} (1 + \omega_{\rm{x}}) \pm Q - \rho_x \frac{2\dot{H}}{H} \right] = \frac{8 \pi G }{3H^2} \rho_{\rm{x}} H \left[ - 3  (1 + \omega_{\rm{x}} - \frac{2\dot{H}}{H^2} \right]  \pm \frac{8 \pi G }{3H^2} Q. \\
\end{split}
\end{gather} 
Where we also have that:
\begin{gather} \label{DSA.4}
\begin{split}
\dot{H}&= \frac{d}{dt} \dot{a}a^{-1}= \frac{\ddot{a}}{a} - \left(\frac{\dot{a}}{a}\right)^2 \quad \rightarrow \quad \frac{\dot{H}}{H^2}= \frac{\ddot{a} a}{a^2} - 1  =-q-1.
\end{split}
\end{gather}
Substituting (\ref{DSA.4}) and $\rho_{\rm{x}} = \frac{3 H^2}{8 \pi G} \Omega_{\rm{x}}$ into (\ref{DSA.3}) gives:
\begin{gather} \label{DSA.5}
\begin{split}
\dot{\Omega}_{\rm{x}} &=  \Omega_{\rm{x}} H \left[ - 3  (1 + \omega_{\rm{x}})  +2q +1 \right]  \pm \frac{8 \pi G }{3H^2} Q = \Omega_{\rm{x}} H \left[ 2q -1 - 3\omega_{\rm{x}} \right]  \pm \frac{8 \pi G }{3H^2} Q. \\
\end{split}
\end{gather}
Substituting in the expression for the deceleration parameter $q$ (\ref{q}) gives:
\begin{gather} \label{DSA.6}
\begin{split}
\dot{\Omega}_{\rm{x}} & = \Omega_{\rm{x}} H \left[ 2\left( \Omega_{\rm{r}}  + \frac{1}{2} \Omega_{\rm{bm}}+ \frac{1}{2} \Omega_{\rm{dm}}  + \frac{1}{2} \Omega_{\rm{de}}  \left(1 +3 \omega_{\rm{de}} \right) \right) -1 - 3\omega_{\rm{x}} \right]  \pm \frac{8 \pi G }{3H^2} Q \\
&= \Omega_{\rm{x}} H \left[ 2 \Omega_{\rm{r}}  + \Omega_{\rm{bm}} + \Omega_{\rm{dm}} + \Omega_{\rm{de}}  \left(1 +3 \omega_{\rm{de}} \right) -1 - 3\omega_{\rm{x}} \right]  \pm \frac{8 \pi G }{3H^2} Q. \\
\end{split}
\end{gather}
This relation holds for either dark matter or dark energy with any coupling function $Q$. If $Q=0$, this reduces back to the same expression for the uncoupled case and may be used not only for dark matter and dark energy but for radiation ($\omega_{\rm{r}}=1/3$) and baryonic matter ($\omega_{\rm{bm}}=0$) as well. For the different components, one therefore has:
\begin{gather} \label{DSA.7}
\begin{split}
\dot{\Omega}_{\rm{de}} &= \Omega_{\rm{de}} H \left[ 2 \Omega_{\rm{r}}  + \Omega_{\rm{bm}}+ \Omega_{\rm{dm}} + \Omega_{\rm{de}}  \left(1 +3 \omega_{\rm{de}} \right) -1 - 3\omega_{\rm{de}} \right]  - \frac{8 \pi G }{3H^2} Q, \\
\dot{\Omega}_{\rm{dm}} &= \Omega_{\rm{dm}} H \left[ 2 \Omega_{\rm{r}}  + \Omega_{\rm{bm}}+ \Omega_{\rm{dm}} + \Omega_{\rm{de}}  \left(1 +3 \omega_{\rm{de}} \right) -1  \right]  + \frac{8 \pi G }{3H^2} Q, \\
\dot{\Omega}_{\rm{bm}} &= \Omega_{\rm{bm}} H \left[ 2 \Omega_{\rm{r}}  + \Omega_{\rm{bm}}+ \Omega_{\rm{dm}} + \Omega_{\rm{de}}  \left(1 +3 \omega_{\rm{de}} \right) -1  \right], \\
\dot{\Omega}_{\rm{r}} &= \Omega_{\rm{r}} H \left[ 2 \Omega_{\rm{r}}  + \Omega_{\rm{bm}}+ \Omega_{\rm{dm}} + \Omega_{\rm{de}}  \left(1 +3 \omega_{\rm{de}} \right) -2 \right].  \\
\end{split}
\end{gather}
Equation (\ref{DSA.6}) reduces back to the $\Lambda$CDM case if $Q=0$ and $\omega_{\rm{de}}=-1$, which can be found in \citep{HobsonTextbook}. For our purposes, we are interested in the parameter space of how dark matter and dark energy evolve with regard to each other. This can be obtained by dividing corresponding dark matter $\dot{\Omega}_{\rm{dm}}$ and dark energy $\dot{\Omega}_{\rm{de}}$ evolution equations (\ref{DSA.7}) by each other, such that: \vspace{0.cm}
\begin{gather} \label{DSA.8}
\begin{split}
\frac{d \Omega_{\rm{de}}}{d \Omega_{\rm{dm}}}&= \frac{\Omega_{\rm{de}} H \left[ 2 \Omega_{\rm{r}}  + \Omega_{\rm{bm}}+ \Omega_{\rm{dm}} + \Omega_{\rm{de}}  \left(1 +3 \omega_{\rm{de}} \right) -1  - 3\omega_{\rm{de}} \right]  -\frac{8 \pi G }{3H^2} Q}{\Omega_{\rm{dm}} H \left[ 2 \Omega_{\rm{r}}  + \Omega_{\rm{bm}}+ \Omega_{\rm{dm}} + \Omega_{\rm{de}}  \left(1 +3 \omega_{\rm{de}} \right) -1\right]  + \frac{8 \pi G }{3H^2} Q}. \\
\end{split}
\end{gather}
This can be used to obtain a set of trajectories or flow lines in the ($\Omega_{\rm{dm}}$, $\Omega_{\rm{de}}$)-plane, which in turn have stable attractor and unstable repulsor points. These will be used to see if the ratio of dark matter to dark energy becomes fixed in the past or present, thus addressing the model's potential to solve the coincidence problem. 
Before considering any IDE models, we will first show how these phase portraits work for the $\Lambda$CDM model, as this will be the standard model to which we will compare our later results.

\section{Two interacting dark energy case studies} \label{sectionQ1}
Now that the general properties of IDE models have been discussed, we will move on to two case studies. This will show that the properties from table \ref{Table_Q_events_relative} hold in general for IDE models of any function $Q$ while highlighting significant differences between the couplings. First, we will consider two of the most common IDE models in the literature, where there is a linear coupling function $Q$ proportional to either the dark matter or dark energy density. These couplings have a strong field theoretical ground \citep{PanField2020} and have the form:
\begin{gather} \label{Q_1_and_2}
\begin{split}
Q = \delta H \rho_{\rm{de}} \quad \quad ; \quad \quad Q = \delta H \rho_{\rm{dm}},
\end{split}
\end{gather}
where $H$ is the Hubble parameter (this dependence on $H$ may naturally arise from first principles) and $\delta$ is a dimensionless coupling constant that determines the strength of the interaction
between dark matter and dark energy \citep{Gavela2009, Gavela2010, Bolotin2015}. It should be noted that the coupling constant $\delta$ is often indicated by $\alpha$ \citep{Vliviita2009, Bhmer2008} (which has an opposite sign to $\delta$) or $\xi$ in the literature 
\citep{ValentinoH02017, ValentinoH02020, ValentinoCU2021, ValentinoCT2020, Lucca2020, Lucca2021, Wang2016}. For these models, we assume that $\delta < -3 \omega$ (so that the coupling strength $|\delta|$ is not too strong \citep{Vliviita2009}). This condition implies $(\delta < -3 \omega ) \rightarrow (\delta + 3 \omega < 0 ).$ Furthermore, since we require that $H>0$ ; $\rho_{\rm{dm}}>0$ ;  $\rho_{\rm{de}}>0$ the sign of $\delta$ will determine the direction of energy flow. Therefore, $\delta>0 \rightarrow Q>0$ corresponds to the iDEDM regime and $\delta<0 \rightarrow Q<0$ to the iDMDE regime. The greatest qualitative difference between the two coupling functions is that $Q_1 \propto \rho_{\rm{dm}}$ and $Q_2 \propto \rho_{\rm{de}}$, which implies that the effect of the coupling will be either most prominent during early dark matter domination, or later dark energy domination respectively. Similarly to what was done in \citep{Gavela2009}, we will use the cosmological parameters of the $\Lambda$CDM model from \cite{Planck2018} for all subsequent figures and calculations, with $\delta=0.25$ for the iDEDM and $\delta=-0.25$ for iDMDE regimes respectively, while temporarily choosing $\omega_{\rm{de}}=\omega=-1$, so that the coupling constant $\delta$ is the only variable that differs from the $\Lambda$CDM model, thus easing comparisons between the $\Lambda$CDM and IDE models. These chosen parameters are purely illustrative. For the latest cosmological parameters obtained from observations for these IDE models, see \citep{NunesS82021, Gariazzo2022}. A preliminary analysis done by us, obtaining cosmological parameters for these IDE models from only supernovae data, can be found in \cite{Marcel2022D, Marcel2022P}.

\subsection{Interaction model: $Q = \delta H \rho_{\rm{de}}$} \label{sectionQ2}
The model $Q= \delta H \rho_{\rm{de}}$ is one of the most common interaction functions in the literature and is more popular than the model $Q= \delta H \rho_{\rm{dm}}$, which we briefly discuss in section \ref{Q1Sum}. A possible explanation for this is that in the iDMDE regime for $Q = \delta H \rho_{\rm{dm}}$ model, $\rho_{\rm{de}}<0$ in the past, as was pointed out in \citep{Gavela2009}. These authors then advocated for a coupling $Q \propto \rho_{\rm{de}}$, since all energy densities remain positive throughout the past universe history, even in the iDMDE ($\delta<0$) regime \citep{Gavela2009}. This result has often been taken at face value in the literature. However, we would like to focus attention on the fact that in the iDMDE ($\delta<0$) regime, these models will \emph{always} suffer from negative dark matter energy densities ($\rho_{\rm{dm}}<0$) during \emph{future} expansion. This can be seen immediately from the interaction function's proportionality to the dark energy density, $Q \propto \rho_{\rm{de}}$. This has an immediate effect on the iDMDE regime. During future expansion, the dark matter density will decrease, and energy will be transferred away from DM to DE until the DM density eventually reaches zero density. However, as there is no mechanism to stop this energy transfer (since energy transfer is only proportional to dark energy density),  the energy transfer will still continue, \textit{inevitably leading to negative dark matter densities ($\rho_{\rm{dm}}<0$) in the future}. This observation should render these models less favourable and should have been noted by many recent papers that have neglected to mention this problem \citep{ValentinoH02020, ValentinoCU2021, ValentinoCT2020, Lucca2020, Lucca2021, NunesS82021, Gariazzo2022}. The exact conditions to ensure positive energy densities throughout the past and future expansion is calculated in section \ref{PECQ2}. However, this result can immediately be seen from the phase portraits of these models.
\subsubsection{Phase portraits}
Assuming the coupling $Q = \delta H \rho_{\rm{de}}= \delta H \left(\frac{3H^2}{8 \pi G }  \Omega_{\rm{de}} \right)$, equation (\ref{DSA.7}) becomes:
\begin{gather} \label{DSA.3.1}
\begin{split}
\frac{d \Omega_{\rm{de}}}{d \Omega_{\rm{dm}}}&= \frac{\Omega_{\rm{de}}  \left[ 2 \Omega_{\rm{r}}  + \Omega_{\rm{bm}}+ \Omega_{\rm{dm}} + \Omega_{\rm{de}}  \left(1 +3 \omega_{\rm{de}} \right) -1  - 3\omega_{\rm{de}}  - \delta \right] }{\Omega_{\rm{dm}}  \left[ 2 \Omega_{\rm{r}}  + \Omega_{\rm{bm}}+ \Omega_{\rm{dm}} + \Omega_{\rm{de}}  \left(1 +3 \omega_{\rm{de}} \right) -1\right] + \delta \Omega_{\rm{de}}    }, \\
\end{split}
\end{gather}
where we have used the fact that $\frac{8 \pi G }{3 H^2} \rho_{\rm{de}}=\Omega_{\rm{de}}$. Since this coupling is not proportional to the dark matter density, it may be seen from (\ref{DSA.7}) that the evolution of baryonic matter and dark matter may be grouped together.
It should be noted that baryonic matter is still separately conserved here and experiences no new interaction. Thus, if dark matter is grouped with baryonic matter and radiation is negligible ($\Omega_r=0$), (\ref{DSA.3.1}) becomes:
\begin{gather} \label{DSA.3.2}
\begin{split}
\frac{d \Omega_{\rm{de}}}{d \Omega_{\rm{m}}}&= \frac{\Omega_{\rm{de}}  \left[ \Omega_{\rm{m}} + \Omega_{\rm{de}}  \left(1 +3 \omega_{\rm{de}} \right) -1  - 3\omega_{\rm{de}}- \delta  \right]  }{\Omega_{\rm{m}}  \left[\Omega_{\rm{m}} + \Omega_{\rm{de}}  \left(1 +3 \omega_{\rm{de}} \right) -1\right] + \delta \Omega_{\rm{de}} }. \\
\end{split}
\end{gather}
Using (\ref{DSA.3.2}), the evolution of matter and dark energy may now be expressed with a phase portrait in the ($\Omega_{\rm{m}}$, $\Omega_{\rm{de}}$)-plane, as seen in figure \ref{PP_Q2}.
\begin{center} 
\begin{figure}[h]
\begin{minipage}{17.5pc}
\includegraphics[width=17.5pc]{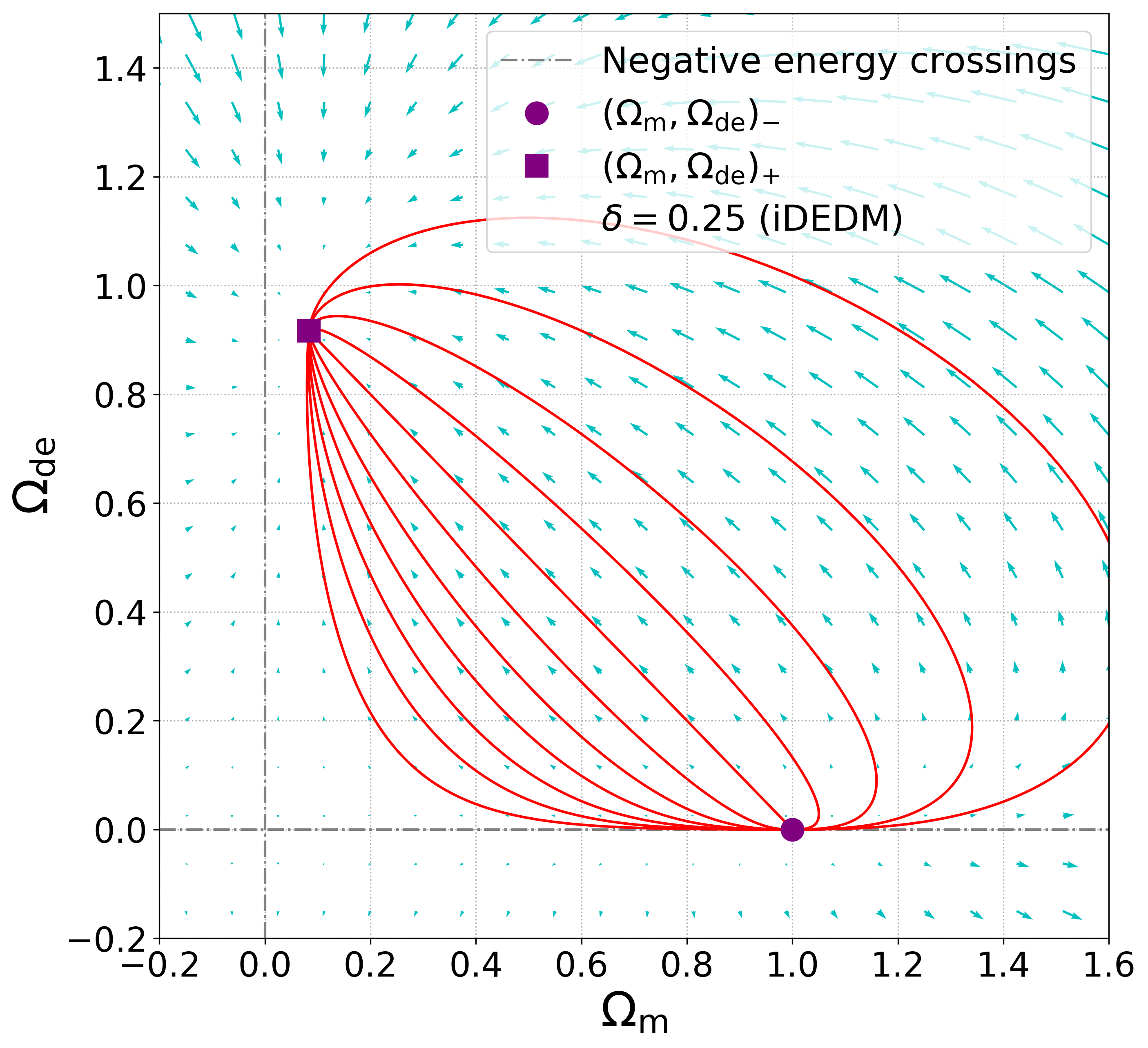}
\end{minipage}\hspace{0pc} 
\begin{minipage}{17.5pc}\vspace{-0.0pc}
\includegraphics[width=17.5pc]{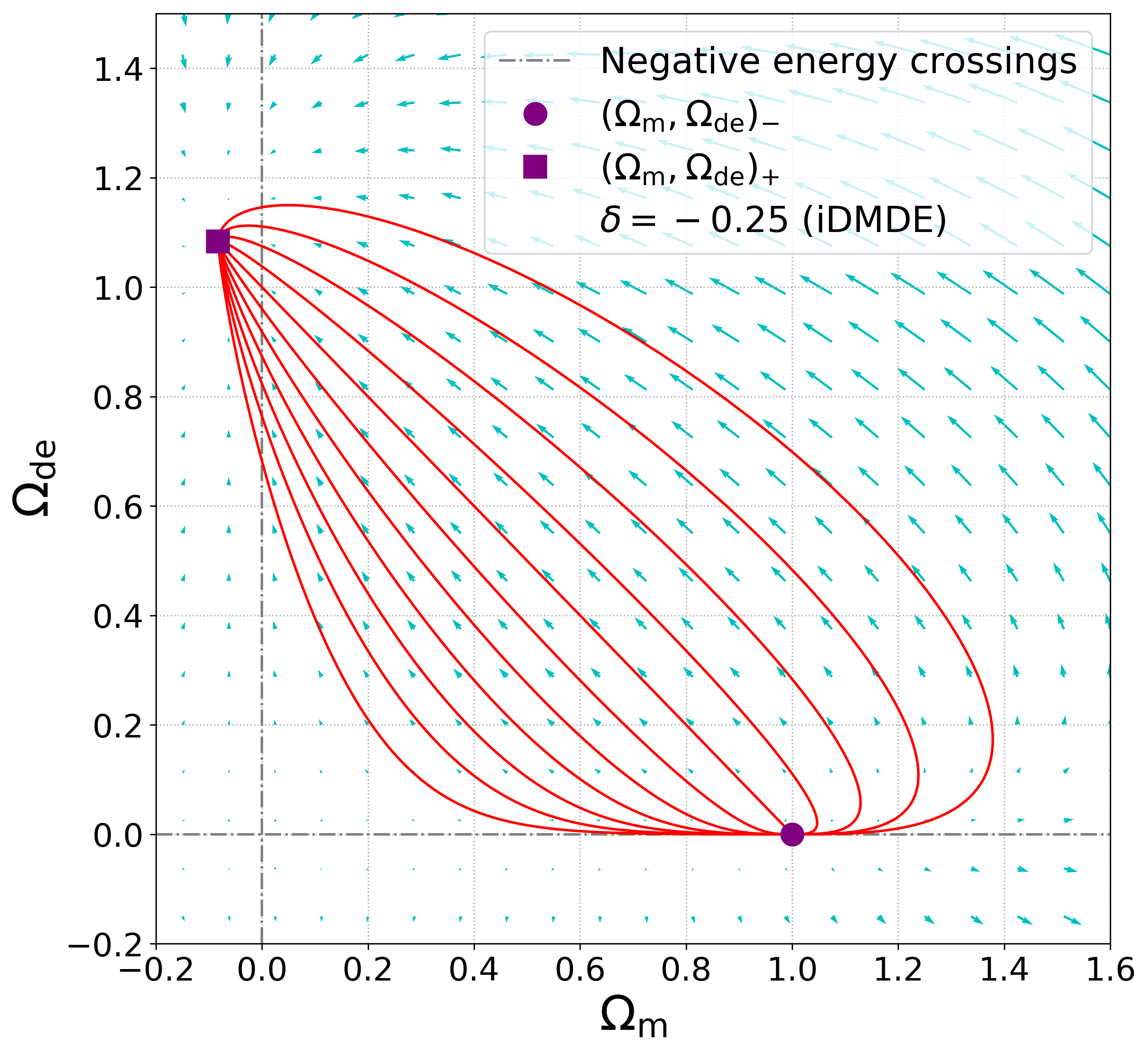}
\end{minipage}
\caption{\label{PP_Q2} Phase portraits for $\Omega_{\rm{dm}}$ and $\Omega_{\rm{de}}$ ($Q = \delta H \rho_{\rm{de}}$)}
\end{figure} \vspace{-1.4pc}
\end{center}
In figure \ref{PP_Q2}, every point on the plane defines a unique trajectory (as indicated by the blue arrows). However, for convenience, we have specified different trajectories (red lines) which pass through specific values for the present matter $\Omega_{\rm(m,0)}=0.3$  and dark energy densities $\Omega_{\rm(de,0)}=0.1,0.2,...,1.1$, as was done in \citep{HobsonTextbook} for the $\Lambda$CDM model. In figure \ref{PP_Q2}, the left panel shows the phase portrait of a positive $\delta$ (iDEDM), while the right panel shows a negative $\delta$ (iDMDE). The equilibrium points are calculated by setting $\dot{\Omega}_{\rm{m}}=0$ and $\dot{\Omega}_{\rm{de}}=0$ in equation (\ref{DSA.3.2}) and solving for $\Omega_{\rm{m}}$ and $\Omega_{\rm{de}}$, yielding:
\begin{gather} \label{DSA.3.3}
\begin{split}
(\Omega_{\rm{m}}, \Omega_{\rm{de}})_{-} = \left( 1,0 \right) \quad \quad ; \quad \quad (\Omega_{\rm{m}}, \Omega_{\rm{de}})_{+}= \left(- \frac{\delta}{ 3 \omega} , 1+\frac{\delta}{ 3 \omega} \right). \\
\end{split}
\end{gather}
Each of the trajectories starts at ($1,0$), which is an unstable repulsor point from which all the trajectories diverge. Finally, these paths all converge again at the stable point $\left(- \frac{\delta}{ 3 \omega}, 1+\frac{\delta}{ 3 \omega} \right)$, which is known as an attractor \citep{HobsonTextbook}. It can be seen that the repulsor point is the same in both cases but that the attractor point is instead shifted by the dark coupling. This highlights the point that the effect of the coupling is more dominant during later dark energy dominance for $Q\propto \rho_{\rm{de}}$. These equilibrium points also highlight the coincidence problem, as the ratio of their coordinates indicates which value $r$ tends to in the past $r_{-}$ or the future $r_{+}$. The $\Lambda$CDM model has $r_-\rightarrow\infty$ in the past, whilst approaching $r_+\rightarrow 0$ in the future. IDE models that can find a constant non-zero or non-infinite value for either $r_-$ or $r_+$ should solve the coincidence problem in either the past or the future, respectively. For this model, we have:
\begin{gather} \label{DSA.3.4}
\begin{split}
r_-= \frac{\Omega_{\rm{(m,-)}}}{ \Omega_{\rm{(de,-)}}} = \frac{1}{0} \rightarrow \infty \quad \quad ; \quad \quad r_+=\frac{\Omega_{\rm{(dm,+)}}}{ \Omega_{\rm{(de,-)}}} \approx \frac{\Omega_{\rm{(m,+)}}}{ \Omega_{(de,+)}}= \frac{- \frac{\delta}{ 3 \omega}}{1+\frac{\delta}{ 3 \omega}} \rightarrow - \frac{\delta }{\delta + 3 \omega}.
\end{split}
\end{gather}
Therefore, this model will not solve the coincidence problem in the past as $r_-$ but will stabilise $r$ in the future $r_+$, thereby solving the coincidence problem for future expansion. Furthermore, positive $\delta>0$ (iDEDM) solves the coincidence problem, but $\delta<0$ (iDMDE) causes negative energy densities. This brings us to the often-overlooked problem of the iDMDE regime for this model. It is clear from (\ref{DSA.3.3}) that $\Omega_{\rm{dm,+}} \approx \Omega_{m,+}= - \frac{\delta}{ 3 \omega}$, alongside ($\omega<0$), must imply that $\delta<$ (iDMDE) leads to a negative energy attractor solution for $\Omega_{\rm{dm}}$. We should also note that baryonic matter is grouped with dark matter. However, in the distant future, it dilutes as in the $\Lambda$CDM model, and its contribution should become negligible, validating the approximation $\Omega_{\rm{dm,+}} \approx \Omega_{\rm{m,+}}$. \\

\subsubsection{Background analytical equations}
To obtain analytical solutions for how the dark matter $\rho_{\rm{dm}}$ and dark energy $\rho_{\rm{de}}$ densities evolve, we need to solve the conservation equations (\ref{IDE1.3}) with $Q=\delta H \rho_{\rm{de}}$, which yields expressions for $\rho_{\rm{dm}}$ and $\rho_{\rm{de}}$:
\begin{align}
\rho_{\rm{dm}} &=  \left( \rho_{\rm{(dm,0)}}    + \rho_{\rm{(de,0)}}  \frac{\delta }{\delta + 3 \omega} \left[ 1  -a^{-\left( \delta + 3 \omega \right) }  \right]\right) a^{-3}, \label{Q2_energy.dm} \\
 \rho_{\rm{de}}  &= \rho_{\rm{(de,0)}} a^{- (\delta + 3 \omega +3) }. \label{Q2_energy.de}
\end{align}
Here (\ref{Q2_energy.dm}) and (\ref{Q2_energy.de}) matching the energy densities found in \citep{Gavela2009, Bolotin2015, Lucca2020,  ValentinoCU2021}. The effective equation of states for this model can be obtained by substituting the coupling equation $Q = \delta H \rho_{\rm{de}}$ into (\ref{3}). The dark matter effective equation of state is then:
\begin{gather} \label{eos.dmQ2}
\begin{split}
\omega^{\rm{eff}}_{\rm{dm}} &= - \frac{Q}{3 H \rho_{\rm{de}}} = - \frac{\delta H \rho_{\rm{de}}}{3 H \rho_{\rm{dm}}} = - \frac{\delta }{3 }   \frac{\rho_{\rm{de}}}{ \rho_{\rm{dm}}}=  - \frac{\delta }{3 }   \frac{1}{r}.
\end{split}
\end{gather}
Similarly, for dark energy, we have the effective equation of state:
\begin{gather} \label{eos.deQ2}
\begin{split}
\omega^{\rm{eff}}_{\rm{de}} &=  \omega + \frac{Q}{3 H \rho_{\rm{de}}} =  \omega + \frac{\delta H \rho_{\rm{de}}}{3 H \rho_{\rm{de}}} =  \omega + \frac{\delta }{3 }.
\end{split}
\end{gather} 
which matches with \citep{Gavela2009, Bolotin2015, Lucca2020, ValentinoCT2020, NunesS82021}. It can be seen that $\omega^{\rm{eff}}_{\rm{dm}}$ is dynamical with a dependence on $r$, while, in contrast $\omega^{\rm{eff}}_{\rm{de}}$ is constant. Equations (\ref{Q2_energy.dm}), (\ref{Q2_energy.de}), (\ref{eos.dmQ2}) and (\ref{eos.deQ2}) also reduce back to the $\Lambda$CDM model when $\delta=0$ and $\omega=-1$. Using the relations $\rho_{\rm{(x,0)}} = \frac{3 H_0^2}{8 \pi G} \Omega_{\rm{(x,0)}}$ and $\Omega_{\rm{x}}  = \frac{8 \pi G}{3 H^2} \rho_{\rm{x}}$, as well as the scalefactor redshift relation $a=(1+z)^{-1}$, we obtain useful equations for the density parameters $\Omega_{\rm{dm}}$ and $\Omega_{\rm{de}}$ from (\ref{Q2_energy.dm}) and (\ref{Q2_energy.de}) that can be added to the standard $\Lambda$CDM model density parameters for baryonic matter $\Omega_{\rm{bm}}$ and radiation $\Omega_{\rm{r}}$: 
\begin{align}
\Omega_{\rm{dm}}   =& \frac{H_0^2}{H^2}  \left( \Omega_{\rm{(dm,0)}}     +  \Omega_{\rm{(de,0)}}  \frac{\delta }{\delta + 3 \omega } \left[ 1  -(1+z)^{ (\delta + 3 \omega) }  \right] \right) (1+z)^{3},  \label{Q2_energy.4} \\
\Omega_{\rm{de}}    =& \frac{H_0^2}{H^2}  \Omega_{\rm{(de,0)}} (1+z)^{ (\delta + 3 \omega +3) },  \label{Q2_energy.5} \\
\Omega_{\rm{bm}}    =& \frac{H_0^2}{H^2}  \Omega_{\rm{(bm,0)}} (1+z)^{3},  \label{Q2_energy.6} \\
\Omega_{\rm{r}}    =& \frac{H_0^2}{H^2}   \Omega_{\rm{(r,0)}} (1+z)^{4} .  \label{Q2_energy.7}
\end{align}

\subsubsection{Positive energy density conditions } \label{PECQ2}
Positive energy conditions for these models were obtained in \citep{Pan2020} from a dynamical systems analysis, where it was claimed that no viable scenarios exist. However, we want to show that viable conditions exist using a similar approach to what was done in \citep{Vliviita2009}. For this model, it can be seen that $\rho_{\rm{de}}$  (\ref{Q2_energy.de}) is always positive (since $a^{- (\delta + 3 \omega +3) } > 0$ for all values of $\delta$), while $\rho_{\rm{dm}}$ (\ref{Q2_energy.dm}) has multiple terms which could become negative.  We now derive the exact conditions to ensure that the $\rho_{\rm{dm}}$ is always positive. To do this, we need to find out where the dark matter energy density crosses the zero energy density boundary and becomes negative so that conditions may be chosen to avoid this zero crossing. This is found when we set the dark matter energy density (\ref{Q2_energy.dm}) equal to zero and solve for $a$:
\begin{gather} \label{Q2_PEC.1}
\begin{split}
 a^{-\left( \delta + 3 \omega \right) }   &= 1 + r_0   \left( \frac{\delta + 3 \omega}{\delta} \right).   \\
\end{split}  	
\end{gather}
Using (\ref{Q2_PEC.1}), we can find solutions where the dark matter energy density crosses zero and becomes negative $(\rho_{\rm{dm}}<0)$. From (\ref{Q2_PEC.1}) and the relation $a=(1+z)^{-1}$, this zero crossing $(\rho_{\rm{dm}}=0)$ happens at exactly the redshift $ z_{\rm{(dm=0)}}$:
\begin{gather} \label{Q2_PEC.2}
\begin{split}
z_{\rm{(dm=0)}} = \left[1 + r_0 \left( \frac{\delta + 3 \omega}{\delta } \right)  \right]^{\frac{1}{\delta +3 \omega}}-1.
\end{split}  	
\end{gather}
Using (\ref{Q2_PEC.1}), we may explore four scenarios, \textbf{(A} - \textbf{(D)}, where the energy density may possibly cross zero and become negative. These scenarios will be either the iDMDE ($\delta < 0$) or iDEDM ($\delta > 0 $) scenarios for either the \textit{past} or the \textit{future}. This leads to:
\begin{gather} \label{Q2_PEC.3}
\begin{split}
                              a^{-\left( \delta + 3 \omega \right) }   &= 1 +r_0  \left( \frac{\delta + 3 \omega}{\delta} \right)  \quad  \quad \text{where} \quad  \quad (\delta + 3 \omega < 0) \\
\end{split}  	
\end{gather} 
\begin{equation} \nonumber
\delta < 0 \Rightarrow
  \begin{cases}
    \text{Past} & (a <1) \quad  \rightarrow \quad \left(   0<  \text{L.H.S.} < 1  \quad; \quad \text{R.H.S.} >1 \right)  \quad  \quad  \textbf{(A)}\\
    \text{Future} & (a >1) \quad  \rightarrow \quad \left(    \text{L.H.S.} > 1 \quad ;\quad \text{R.H.S.} >1 \right)   \quad   \quad \quad \; \; \;  \textbf{(B)}\\
  \end{cases}     
\end{equation}
\begin{equation} \nonumber
\delta > 0 \Rightarrow
  \begin{cases}
    \text{Past} & (a <1) \quad  \rightarrow \quad\left( 0< \text{L.H.S.} < 1  \quad; \quad \text{R.H.S.} <1 \right)  \quad  \quad  \textbf{(C)}\\
    \text{Future} & (a >1) \quad  \rightarrow \quad \left( \text{L.H.S.} > 1 \quad ;\quad \text{R.H.S.} <1 \right)   \quad  \quad \quad \; \; \; \textbf{(D)}\\
  \end{cases}     
\end{equation}\\
Here we can immediately see that for both $\textbf{(A)}$ and $\textbf{(D)}$ the L.H.S. and R.H.S. will never cross, which means that there will be no solution for (\ref{Q2_PEC.3}) and thus the $\rho_{\rm{dm}}$ will never cross zero and become negative. Therefore,  $\rho_{\rm{dm}}$ will always remain positive for scenerio's $\textbf{(A)}$ (\textit{Past} expansion with $\delta<0$) and $\textbf{(D)}$ (\textit{future} expansion with $\delta>0$). Furthermore, scenario \textbf{(B)} will always have a solution, and therefore, the dark energy density will always become negative in the \textit{future}, as shown by the attractor point in figure {\ref{PP_Q2}}. 
\begin{gather} \label{Q2_PEC.4}
\begin{split}
  1 + r_0 \left( \frac{\delta + 3 \omega}{\delta } \right) &< 0 
         \quad \quad           \rightarrow    \quad \quad       \delta  < -  \frac{3 \omega}{\left( 1 + \frac{1}{r_0} \right) }.  
\end{split}  	
\end{gather}
Thus, if condition (\ref{Q2_PEC.4}) is met, then scenario \textbf{(C)} (\textit{Past} expansion with $\delta>0$) will always have positive energy densities. 
Therefore, since both \textbf{(C)} and \textbf{(D)} will always have positive energy densities, the positive coupling $\delta>0$ (with condition (\ref{Q2_PEC.4}) met) may be seen as physical. Since the condition (\ref{Q2_PEC.4}) holds, it implies that the condition $\delta < -3 \omega$ must necessarily hold as well. Taking the conditions $ \left( \delta > 0 \right); \left( \delta < -3 \omega \right)$ and $\left( \delta  < -  \frac{3 \omega}{\left( 1 + 1/r_0 \right) } \right) $ together, a general condition is obtained to ensure positive energy densities for this model:
\begin{gather} \label{Q2_PEC.5}
\begin{split}
     0 <   \delta  < -  \frac{3 \omega}{\left( 1 + \frac{1}{r_0} \right) }.  \\
\end{split}  	
\end{gather}
The energy densities for all these conditions may be encapsulated in table \ref{Tab_PEC_Q2} below.
\begin{table}[h]
\centering
\begin{tabular}{|c|c|c|c|c|c|}
\hline
Conditions & $\rho_{\rm{dm}}$ (Past) & $\rho_{\rm{dm}}$ (Future) &  $\rho_{\rm{de}}$ (Past) & $\rho_{\rm{de}}$ (Future) & Physical\\
\hline  \hline
\rule{0pt}{12pt}     $ 0 <   \delta  < -  \frac{3 \omega}{\left( 1 + \frac{1}{r_0} \right) }$& +& +  & + & + & $\surd$ \\
\hline
\rule{0pt}{12pt} $\delta > 0 $ ; $\delta  >-  \frac{3 \omega}{\left( 1 + \frac{1}{r_0} \right) }$ & $- $& + & + & + & X \\
\hline
\rule{0pt}{12pt} $\delta < 0$ & + & - & + & + & X \\
\hline
\end{tabular}
\caption{Conditions for positive energy densities throughout cosmic evolution $\left( Q = \delta H \rho_{\rm{de}} \right)$}
 \label{Tab_PEC_Q2}
\end{table} \\
In table \ref{Tab_PEC_Q2}, $(+)$ means that the energy densities will always remain positive, $(-)$ means that the energy densities will always become negative somewhere in the cosmic evolution. Any scenario leading to negative energy densities should be considered \textit{unphysical}. Thus, only systems that abide by the condition (\ref{Q2_PEC.5}) may be considered \textit{physical}.
From this condition  (\ref{Q2_PEC.5}), it may be concluded that only IDE models where energy flows from dark energy to dark matter iDEDM ($\delta>0$) should be seriously considered as couplings where energy flows from dark matter to dark energy iDMDE ($\delta<0$) will always lead to either negative energies in the past or the future at redshift $ z_{\rm{(dm=0)}}$  (\ref{Q2_PEC.2}). This only holds for the coupling $Q = \delta H \rho_{\rm{de}}$, and may not be the case for other coupling models.\\

\subsubsection{Cosmic coincidence problem } \label{CCPQ2}
For this model, the coincidence problem is not solved in the past but instead in the future. This can be seen from the repulsor point $r_-=\infty$ (\ref{DSA.3.4}) being the same as in the $\Lambda$CDM model. Conversely, the attractor point $r_+$ in (\ref{DSA.3.4}) shows that this model should solve the coincidence problem in the future, at least for the iDEDM regime. We will now reproduce these results from the analytical expression for $\rho_{\rm{dm}}$ (\ref{Q2_energy.dm}) and $\rho_{\rm{de}}$ (\ref{Q2_energy.de}), which we may use to obtain the following simplified expression for $r$ in terms of redshift $z$:
\begin{gather} \label{r.2_Q2}
\begin{split}
r (z) = \frac{\rho_{\rm{dm}}(z)}{\rho_{\rm{de}} (z)} =  \left(r_0  + \frac{\delta }{\delta + 3 \omega} \right)  (1+z)^{ -(\delta + 3 \omega)}  - \frac{\delta }{\delta + 3 \omega}\\
\end{split}
\end{gather}
From (\ref{r.2_Q2}), it can be seen that $r$ has the proportionality, such that:
\begin{gather} \label{r.3_Q2}
\begin{split}
r \propto   a^{ (\delta + 3 \omega)}  \quad \quad \rightarrow \quad \quad \zeta_{\rm{Q_1}} = \zeta_{\rm{Q}} = -3\omega -\delta.
\\
\end{split}
\end{gather}
For the $\Lambda$CDM model $\zeta_{\Lambda\rm{CDM}}=3$, and for a general uncoupled model $\zeta=-3 \omega$, thus from (\ref{r.3_Q2}) it can be seen that:
\begin{equation} \label{r.4_Q2}
\zeta_{\rm{Q}} = -3\omega -\delta \rightarrow
  \begin{cases}
 \text{if }\delta >0 \text{ (iDEDM)} \quad \rightarrow \quad \zeta_{\rm{Q}}< \zeta & \text{\textit{alleviates} coincidence problem}  \\
   \text{if } \delta <0 \text{ (iDMDE)} \quad \rightarrow \quad \zeta_{\rm{Q}}> \zeta & \text{\textit{worsens} coincidence problem}.  \\
  \end{cases}     
\end{equation}
This behaviour coincides with the original analysis in (\ref{CCP.5}). Furthermore, this effect becomes more extreme in both the distant past (at large redshifts $(1+z)\rightarrow \infty$) and the distant future (at redshifts $(1+z)\rightarrow 0$). This can be seen by considering these limits for (\ref{r.2_Q1}), while noting the condition $\delta + 3 \omega <0$, thus:
\begin{gather} \label{r.5_Q2}
\begin{split}
\lim_{(1+z)\to \infty} r_- \rightarrow  \infty,  \quad \quad ; \quad \quad  \lim_{(1+z)\to 0} r_+  &= \rightarrow - \frac{\delta }{\delta + 3 \omega} .  \\
\end{split}
\end{gather}
These results match what was found from the phase portrait in figure \ref{PP_Q2}, with the repulsor point $r_-$ and attractor point $r_+$ (\ref{DSA.3.4}) being the same as the $(1+z)\rightarrow \infty$  and $(1+z)\rightarrow 0$ redshift limits found for $r$ in (\ref{r.5_Q2}), respectively.
Furthermore, in the distant future, $r$ has the proportionality:
\begin{gather} \label{r.6_Q2}
\begin{split}
\lim_{(1+z)\to 0} r_+ \propto   a^{0} \quad
\rightarrow \quad \zeta_{\rm{(Q,-)}} = 0.
\\
\end{split}
\end{gather}
Since $r$ is constant and $\zeta_{\rm{(Q,+)}}=0$, this model solves the coincidence problem for future expansion. This only holds for the $\delta>0$ (iDEDM) regime, since $\delta<0$ (iDMDE) will lead to a negative constant $r_+$ due to $\rho_{\rm{dm}}$ which becomes negative at $z_{\rm{(dm=0)}}$ (\ref{Q2_PEC.2}), as shown in table \ref{Tab_PEC_Q2}, which is unphysical.
Thus, for $(1+z)\to 0$ in the future, we have:
\begin{equation} \label{r.7_Q2}
\lim_{(1+z)\to 0} \zeta_{\rm{Q}} = 0
  \begin{cases}
 \text{if }\delta >0 \rightarrow r_-= +constant & \text{\textit{solves} coincidence problem}  \\
   \text{if } \delta <0 \rightarrow r_-= -constant & \text{negative energy densities (unphysical)}.  \\
  \end{cases}     
\end{equation}
To understand this behaviour, we can consider how the effective equations of state $\omega^{\rm{eff}}$ for this model evolve. To do this, we first need the explicit relation for $\omega^{\rm{eff}}_{\rm{dm}}$, which is obtained by substituting in $r$ from (\ref{r.2_Q2}) into (\ref{eos.dmQ2}):
\begin{gather} \label{eos.3Q2}
\begin{split}
\omega^{\rm{eff}}_{\rm{dm}} &= - \frac{\delta }{3 } \frac{1}{r} = - \frac{\delta }{3 }   \frac{1}{ \left(r_0  + \frac{\delta }{\delta + 3 \omega} \right)  (1+z)^{ -(\delta + 3 \omega)}  - \frac{\delta }{\delta + 3 \omega} }.
\end{split}
\end{gather}
In the distant past the ratio $r_- \rightarrow \infty$, while in the distant future $r_+ \rightarrow - \frac{\delta}{\delta + 3 \omega}$, as was independently shown in both (\ref{DSA.3.4}) and  (\ref{r.5_Q2}). Noting that $\omega^{\rm{eff}}_{\rm{de}} = \omega + \frac{\delta }{3 }$ from (\ref{eos.deQ2}), we can see how the dynamical effective equation of state $\omega^{\rm{eff}}_{\rm{dm}}$ behaves in both the distant past and future:  
\begin{gather} \label{eos.4Q2}
\begin{split}
\omega^{\rm{eff}}_{\rm{dm}} &=  - \frac{\delta}{3 } \frac{1}{r}
  \begin{cases}
  \text{Distant past } \quad \left(r = r_{-}\right)  \text{: }& \quad\omega^{\rm{eff}}_{\rm{dm}}  =  - \frac{\delta }{3 } \frac{1}{\infty} = 0 = \omega_{\rm{dm}}  \\
\text{Distant future } \left( r=r_+ \right) \text{: }& \quad \omega^{\rm{eff}}_{\rm{dm}} =    - \frac{\delta }{3 } \left(  \frac{\delta }{\delta + 3 \omega}\right)=  \omega + \frac{\delta }{3 } = \omega^{\rm{eff}}_{\rm{de}}.
  \end{cases}.    
\end{split}
\end{gather}
The effective equations of state for dark matter and dark energy are, therefore, the same in the distant future ($ \omega^{\rm{eff}}_{\rm{dm}}= \omega^{\rm{eff}}_{\rm{de}}$). This shows that dark matter and dark energy redshift and dilute at the same rate in the future, effectively solving the coincidence problem by keeping the ratio of dark matter to dark energy constant. This shows that whenever $r=+constant\rightarrow \zeta=0$,  we also have $\omega^{\rm{eff}}_{\rm{dm}}= \omega^{\rm{eff}}_{\rm{de}}$. \\
Furthermore, we can also see that in the distant past, $ \omega^{\rm{eff}}_{\rm{dm}}= \omega_{\rm{dm}}$. The effect of the coupling on dark matter will thus become negligible for past expansion, effectively mimicking the behaviour of uncoupled dark matter. These predictions agree with what was found by \citep{delCampo2009, Wang2016} and is confirmed by plotting $r$ (\ref{r.2_Q2}) alongside both $\omega^{\rm{eff}}_{\rm{dm}}$ (\ref{eos.3Q2}) and $\omega^{\rm{eff}}_{\rm{de}}$ (\ref{eos.deQ2}) in figure \ref{Coincidence_Problem_Q2}. 
\begin{center}\vspace{-0.2pc}
\begin{figure}[h]
\begin{minipage}{18.1pc}
\includegraphics[width=18.1pc]{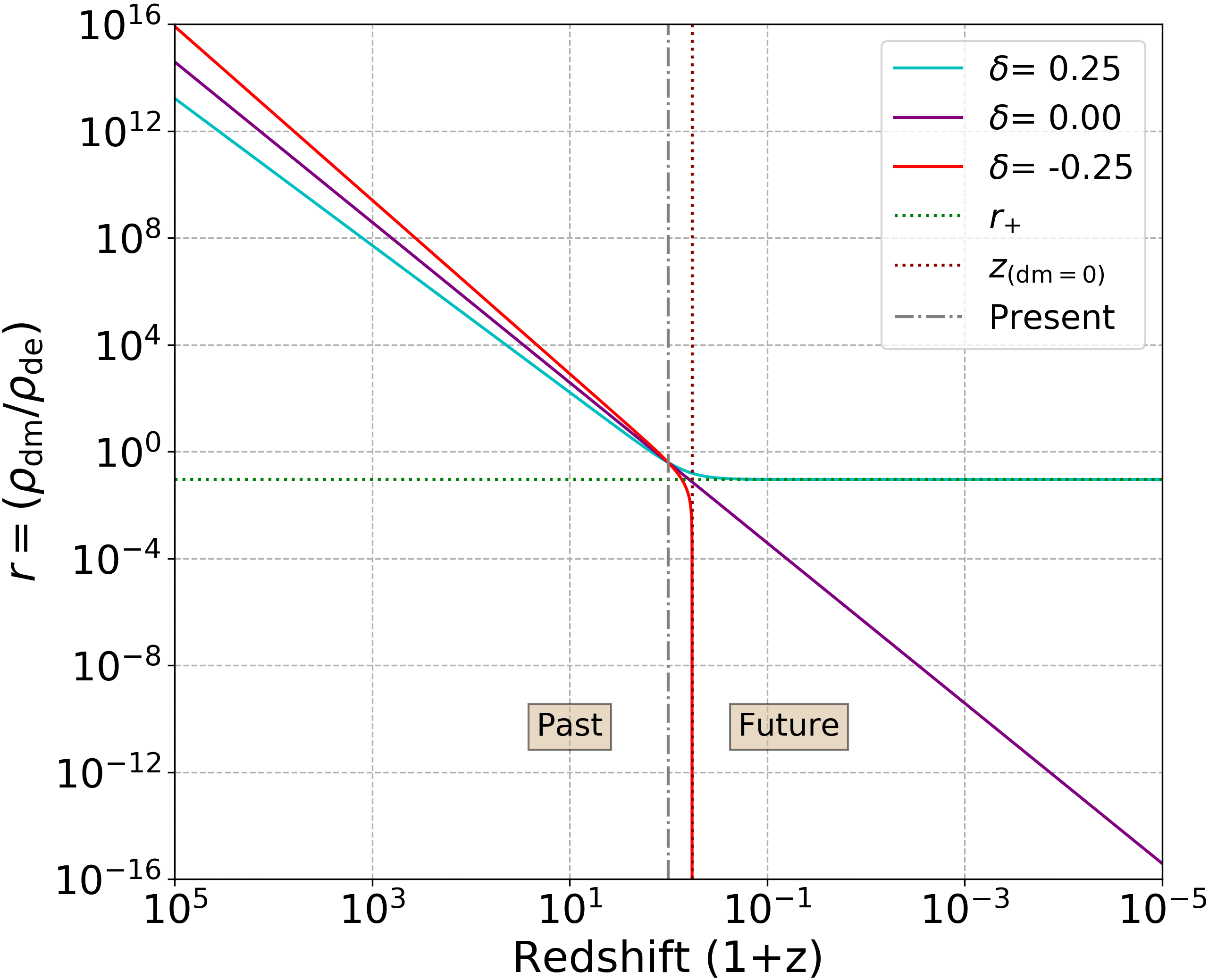}
\end{minipage}
\hspace{-0.25pc} 
\begin{minipage}{17.3pc}\vspace{-0.0pc}
\includegraphics[width=17.3pc]{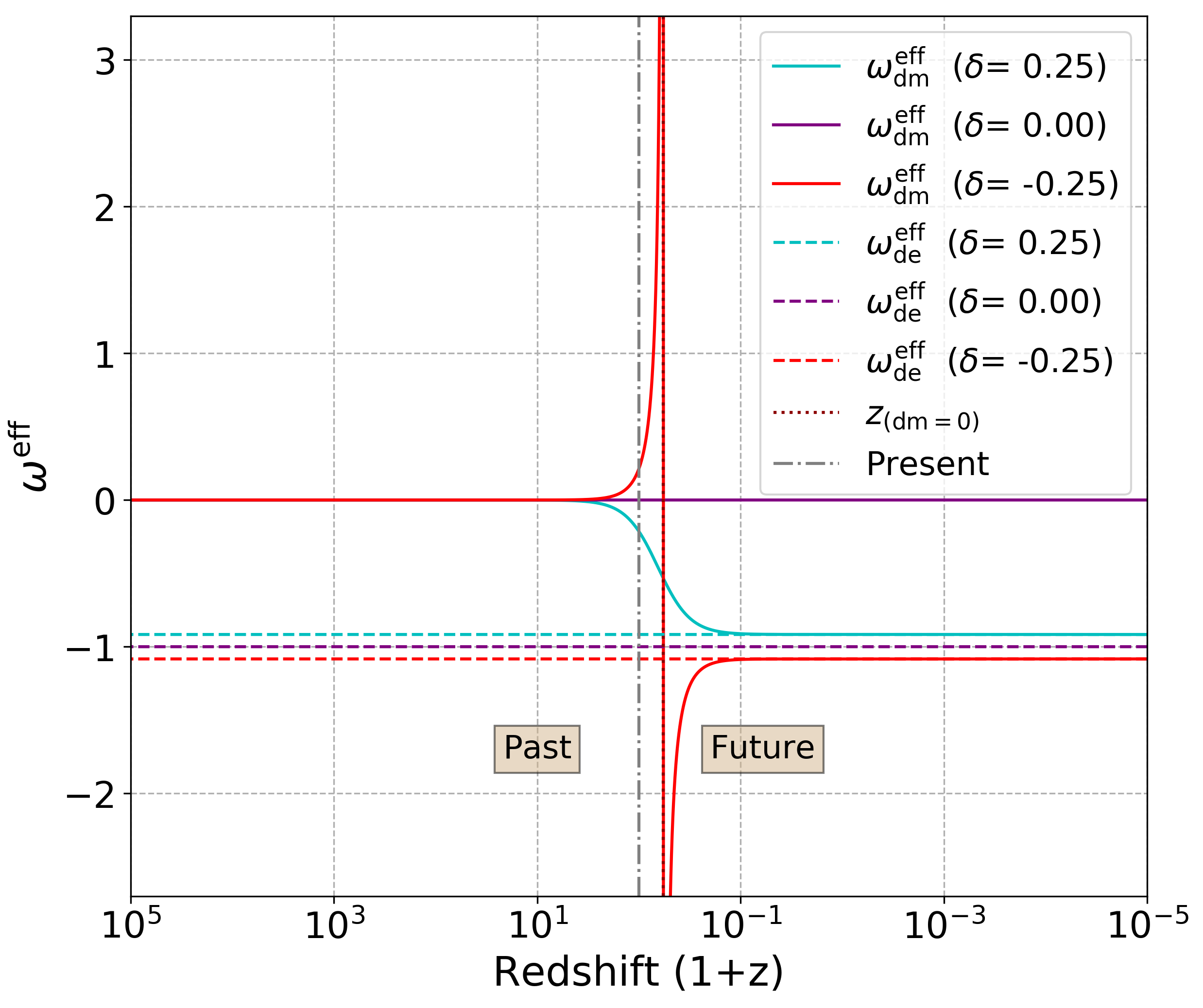}
\end{minipage}
\caption{\label{Coincidence_Problem_Q2} Coincidence problem and effective equations of state ($Q = \delta H \rho_{\rm{de}}$)}
\end{figure} \vspace{-1.8pc}
\end{center}
From the left panel in figure \ref{Coincidence_Problem_Q2}, it can be seen that for the coupled model with $\delta>0$ (iDEDM), $r$ differs with many orders of magnitude in the past but converges to a constant value in the future $r\rightarrow r_{+}$ (indicated by the dashed green line), as predicted by (\ref{DSA.3.4}) and (\ref{r.5_Q2}), making the present value less coincidental. The coincidence problem is thus solved for the future expansion history. This coincides with the right panel where $\omega^{\rm{eff}}_{\rm{dm}}= \omega^{\rm{eff}}_{\rm{de}}$, as shown in (\ref{eos.4Q2}). The coincidence problem is also alleviated for the past expansion since the slope of $r$ is smaller (as predicted by (\ref{r.4_Q2})), which coincides with $\omega^{\rm{eff}}_{\rm{de}} > \omega_{\rm{de}}$ from (\ref{IDE1.3}), causing a smaller difference in ($\omega^{\rm{eff}}_{\rm{dm}}-\omega^{\rm{eff}}_{\rm{de}}$).\\ 
Conversely, for $\delta<0$ (iDEDM), we have $\omega^{\rm{eff}}_{\rm{de}} < \omega_{\rm{de}}$, which worsens the coincidence problem for the past expansion history (since the slope is greater than the case $\delta = 0$). For future expansion, it can also be seen that $r$ becomes zero, while $\omega^{\rm{eff}}_{\rm{de}}$ diverges at the same point. This is due to the dark matter density $\rho_{\rm{dm}}$ which becomes zero in the future at redshift $z_{\rm{(dm=0)}}$ (red dotted line) from (\ref{Q2_PEC.2}), and then stays negative for the rest of the future expansion, indicating the unviability of the iDMDE regime. Thus, the results from (\ref{r.5_Q2}), (\ref{r.4_Q2}), (\ref{r.7_Q2}) and (\ref{eos.4Q2}) can clearly be seen in figure \ref{Coincidence_Problem_Q2} and may be summarised as:\vspace{-0.25pc} \\ \vspace{-0.1pc}
\begin{equation} \label{eos.summary_Q2} 
\delta >0 \; (\text{iDEDM})
  \begin{cases}
 \text{Past expansion: }  \quad \quad   \omega^{\rm{eff}}_{\rm{de}} > \omega_{\rm{de}} \; (\zeta_{\rm{Q}}< \zeta) & \text{\textit{alleviates} coincidence problem} \quad \quad \quad \quad \\
 \text{Future expansion: } \quad \omega^{\rm{eff}}_{\rm{dm}} = \omega^{\rm{eff}}_{\rm{de}} \; (\zeta_{\rm{Q}}=0) & \text{\textit{solves} coincidence problem,}
     \end{cases}     
\end{equation}
\vspace{-0.4cm} \hspace{-0.2cm} \begin{equation} \nonumber 
  \delta <0  \; (\text{iDMDE})
  \begin{cases}
 \text{Past expansion: } \quad  \quad    \omega^{\rm{eff}}_{\rm{de}}<\omega_{\rm{de}} \; (\zeta_{\rm{Q}}> \zeta) & \text{\textit{worsens} coincidence problem}   \\
 \text{Future expansion: }  \quad \omega^{\rm{eff}}_{\rm{dm}} = \omega^{\rm{eff}}_{\rm{de}} \;(\rho_{\rm{de}}<0) &  \text{negative energy densities.}
  \end{cases}       \quad \quad \; \;
\end{equation} \vspace{-0.0cm}\\

\subsubsection{Evolution of energy densities and cosmic equalities } \label{EnergyQ2}
\begin{figure}[h] 
 \centering 
 \includegraphics[width=13cm, height=6.5 cm]{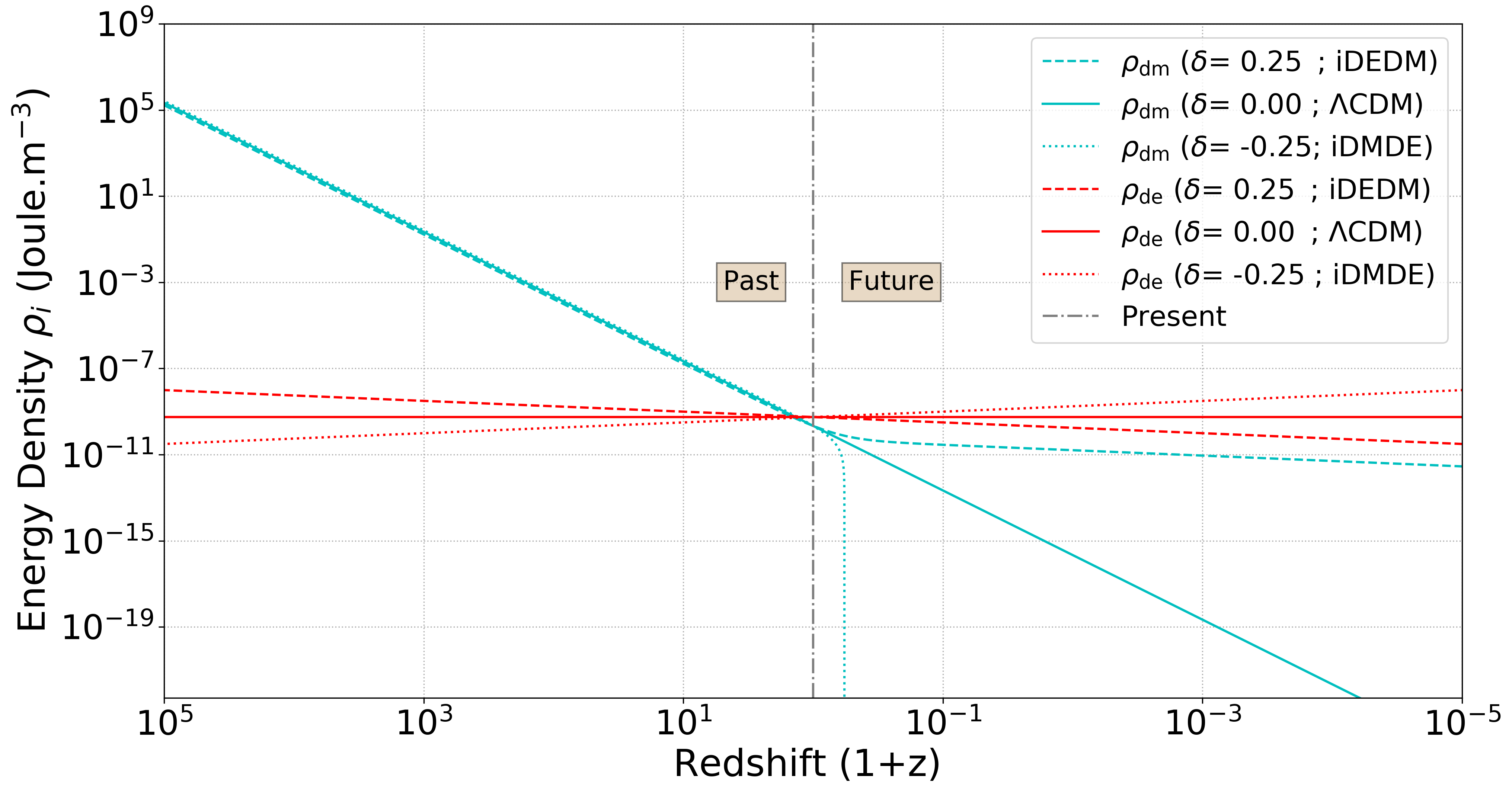}
\vspace{-0.3cm} \caption{Energy densities $\rho$ vs redshift - ($Q = \delta H \rho_{\rm{de}}$)}
\label{fig_rho_Q2}
\end{figure}

\begin{figure}[h]\vspace{-0.1cm}
 \centering
 \includegraphics[width=15.5cm, height=8 cm]{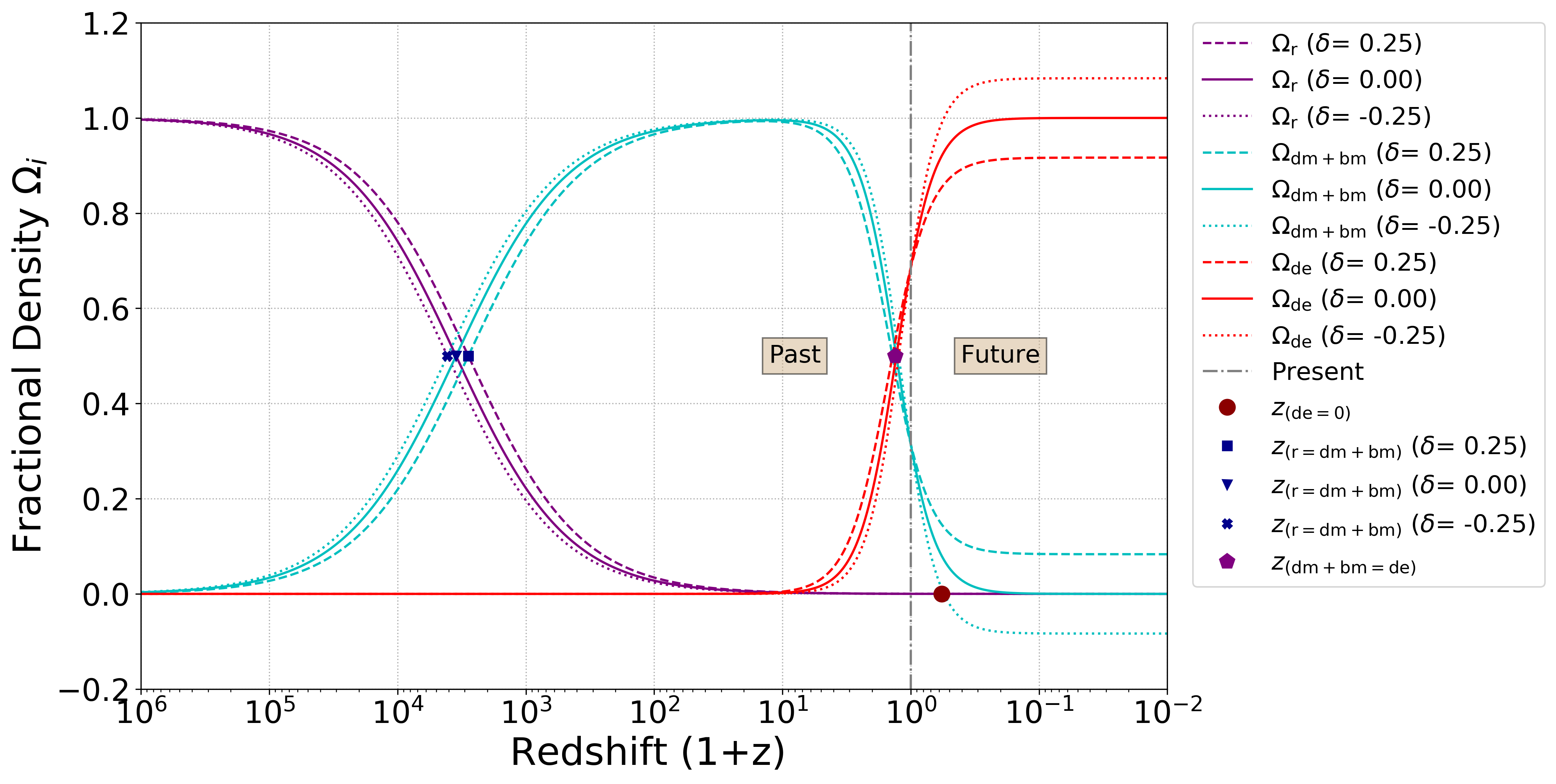} \vspace{-0.7cm}
 \caption{Density parameters vs redshift - ($Q = \delta H \rho_{\rm{de}}$)}
\label{fig_Omega_Q2}  \vspace{-0.2cm}
\end{figure}

For this model, we saw that the coincidence problem is solved for the future (\ref{eos.summary_Q2}). This can be clearly seen by plotting $ \rho_{\rm{dm}}$ (\ref{Q2_energy.dm}) and $\rho_{\rm{de}}$ (\ref{Q2_energy.de}) against redshift $z$ in figure \ref{fig_rho_Q2}.
 \\
In figure \ref{fig_rho_Q2}, we see that for $\delta >0$ (iDEDM), dark matter receives energy from dark energy, causing  $\rho_{\rm{dm}}$ to redshift slower $\omega^{\rm{eff}}_{\rm{dm}} < \omega_{\rm{dm}}$  (smaller slope), while  $\rho_{\rm{de}}$ redshifts faster (greater slope). This behaviour alleviates the coincidence problem in the past. In the future the slope at which $\rho_{\rm{dm}}$ and $\rho_{\rm{de}}$ redshift becomes the same, coinciding with $\omega^{\rm{eff}}_{\rm{dm}} = \omega^{\rm{eff}}_{\rm{de}}$ (\ref{eos.4Q2}) and the coincidence problem being solved, while $ \rho_{\rm{dm}}$ dilutes similar to the $\Lambda$CDM model in the past where $\omega^{\rm{eff}}_{\rm{dm}} = \omega_{\rm{dm}}$ (\ref{eos.4Q2}). All these observations coincide with (\ref{eos.summary_Q2}). \\
It may also be noted that if $\delta >0$ (iDEDM), $\rho_{\rm{de}}$ decreases over time; while if $\delta <0$ (iDMDE), $\rho_{\rm{de}}$ increases over time. The dark energy, therefore, effectively behaves like either quintessence or phantom dark energy, respectively, with an equation of state $\omega^{\rm{eff}}_{\rm{de}}=\omega + \frac{\delta }{3}$. Since this effect continues indefinitely, it may cause a big rip singularity in the future at the time (\ref{BR.5_Q2}).
\\
We can now show that other implications of a dark coupling from table \ref{Table_Q_events_relative} hold for this coupling function. This is done by plotting the density parameters of dark matter $\Omega_{\rm{dm}}$ (\ref{Q2_energy.4}), dark energy $\Omega_{\rm{de}}$ (\ref{Q2_energy.5}), radiation $\Omega_{\rm{r}}$ (\ref{Q2_energy.7}) and baryonic matter $\Omega_{\rm{bm}}$ (\ref{Q2_energy.6}) in figure \ref{fig_Omega_Q2}, as was done in figure 2 of \citep{Gavela2009} for \emph{only the past expansion}, but we include the crucial future expansion as well.
For $\delta >0$ (iDEDM), there will also be a time in the future when $\rho_{\rm{dm}}=0$ at redshift $z_{\rm{(dm=0)}}$ (\ref{Q2_PEC.2}), after which $\rho_{\rm{dm}}<0$ for the rest of expansion, which is unphysical. The predicted value for  $z_{\rm{(dm=0)}}$ is indicated by the red marker in figure \ref{fig_Omega_Q2}.
  \\ 
From figures \ref{fig_rho_Q2} and \ref{fig_Omega_Q2}, it is seen that for $\delta > 0$ (iDEDM), there is \textit{less} dark matter and more dark energy in the past, and vice versa for $\delta < 0$ (iDMDE). For $\delta > 0$ the matter-radiation equality happens \textit{later} and the matter-dark energy equality \textit{earlier} in cosmic history, with the opposite holding for $\delta < 0$ \citep{Gavela2009}. Analytical expressions giving the exact redshift where the radiation-matter $z_{(\rm{r=dm+bm})}$ (\ref{EQ_Q2_3}) equality occurs may be calculated for this model by setting $\Omega_{\rm{(bm,0)}}+\Omega_{\rm{(dm,0)}}=\Omega_{\rm{(r,0)}}$ from equations (\ref{Q2_energy.4}), (\ref{Q2_energy.6}) and (\ref{Q2_energy.7}) and solving for $z$ (using an approximation neglecting a small term), giving:
\begin{gather} \label{EQ_Q2_3}
\begin{split}
z_{(\rm{r=dm+bm})}&\approx \left( \frac{\Omega_{\rm{(bm,0)}} +  \Omega_{\rm{(dm,0)}}     + \Omega_{\rm{(de,0)}}  \frac{\delta }{\delta + 3 \omega}}{ {\Omega_{\rm{(r,0)}}}} \right) -1 . \\
\end{split}
\end{gather}
The matter dark energy equality $z_{(\rm{dm+bm=de})}$ occurs when $\Omega_{\rm{(bm,0)}}+\Omega_{\rm{(dm,0)}}=\Omega_{\rm{(de,0)}}$ from equations (\ref{Q2_energy.4}), (\ref{Q2_energy.5}) and (\ref{Q2_energy.6}) and solving for $z$, giving:
\begin{gather} \label{EQ_Q2_6}
\begin{split}
   z_{(\rm{dm+bm=de})} &=        \left( \frac{\frac{\Omega_{\rm{(bm,0)}} +  \Omega_{\rm{(dm,0)}}}{\Omega_{\rm{(de,0)}}} +  \frac{\delta }{\delta + 3 \omega}}{\left(1+  \frac{\delta }{\delta + 3 \omega} \right)} \right)^{\left(\frac{1}{ \delta + 3 \omega } \right)}
-1 .
\end{split}
\end{gather}
Equation (\ref{EQ_Q2_3}) and (\ref{EQ_Q2_10}) was analytically solved, with the results shown in tables \ref{Table_Q2_iDEDM_events} and \ref{Table_Q2_iDMDE_events}. These results are indicated by the markers in figure \ref{fig_Omega_Q2}, matching with where the corresponding densities intersect. From these results, we confirm what was shown in table \ref{Table_Q_events_relative}:\\
\vspace{-0.0cm}\begin{equation} \label{eq_Q2}
\delta >0 \; (\text{iDEDM})
\begin{cases}
\text{Radiation-matter equality: }  \quad \quad z_{\rm{IDE}}<z_{\Lambda\rm{CDM}}  & \text{happens \textit{later} than }\Lambda\text{CDM} \\
\text{Matter-dark energy equality: } \quad z_{\rm{IDE}}>z_{\Lambda\rm{CDM}}  & \text{happens \textit{earlier} than }\Lambda\text{CDM}, \quad \quad \quad \quad
\end{cases}   
\end{equation}
\vspace{-0.5cm} \hspace{-0.2cm}
\begin{equation}  \nonumber
\delta <0  \; (\text{iDMDE})
\begin{cases}
\text{Radiation-matter equality: }  \quad \quad z_{\rm{IDE}}>z_{\Lambda\rm{CDM}}  & \text{happens \textit{earlier} than }\Lambda\text{CDM} \\
\text{Matter-dark energy equality: } \quad z_{\rm{IDE}}<z_{\Lambda\rm{CDM}}  & \text{happens \textit {later} than }\Lambda\text{CDM}, \quad  \quad \quad \quad \quad
   \end{cases}
\end{equation}\\
In figure \ref{fig_rho_Q2} it may also be seen there is complete matter domination $(\Omega_{\rm{dm}}, \Omega_{\rm{de}})_{-} = \left(1,0 \right)$, as in the $\Lambda$CDM case. Conversely, dark energy never completely dominates in the future, but instead, dark matter and dark energy have the density parameters $(\Omega_{\rm{{dm+bm}}}, \Omega_{\rm{de}})_{+}= \left(- \frac{\delta}{ 3 \omega}, 1+\frac{\delta}{ 3 \omega} \right)$ from the attractor point (\ref{DSA.3.3}).\\
\subsubsection{Evolution of deceleration parameter}
\begin{center}
\begin{figure}[h]
\begin{minipage}{17.6pc}
\includegraphics[width=17.6pc]{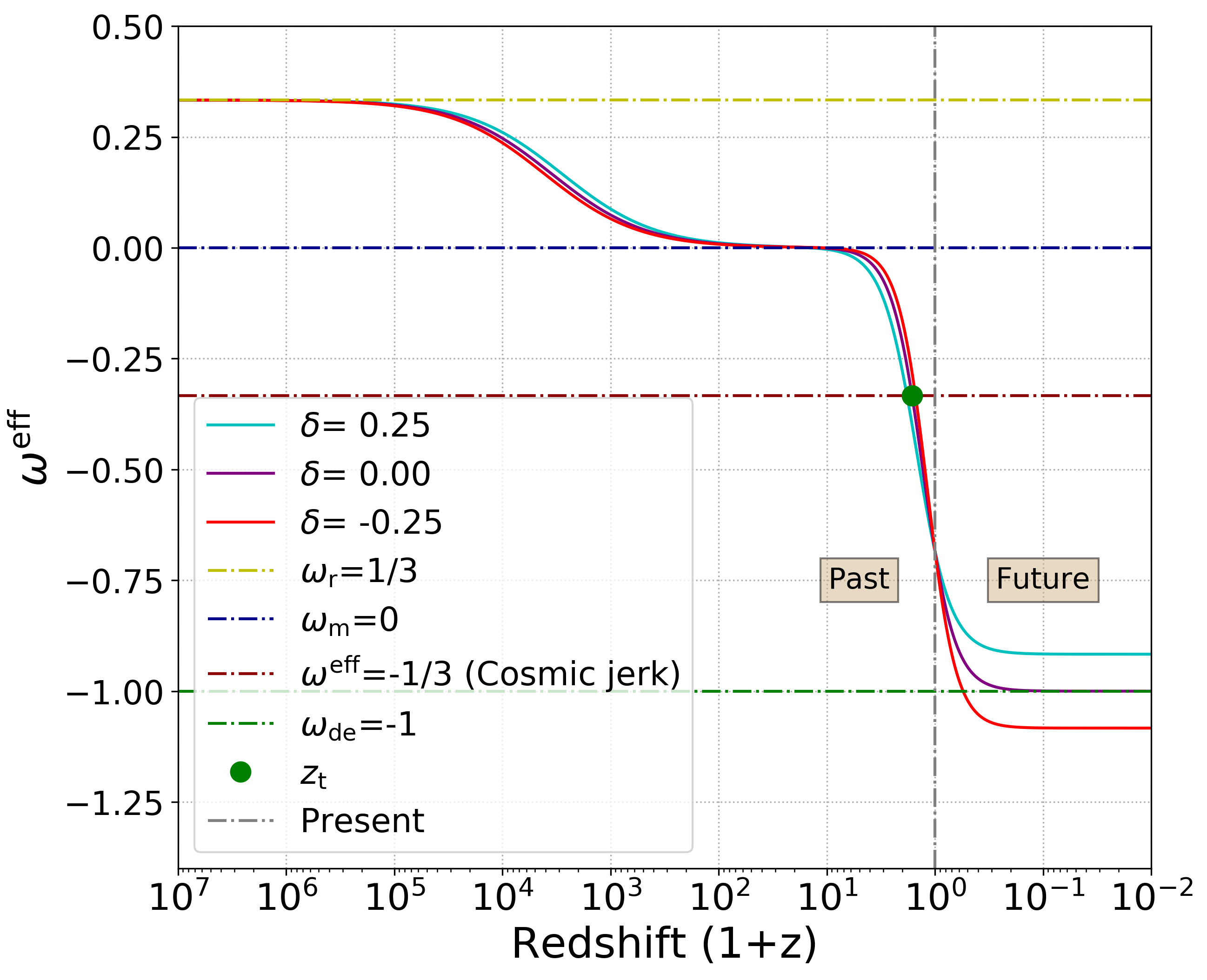} \vspace{-1.8pc}
\caption{\label{omega_eff_evo_Q2} Evolution of effective equation of state $\omega^{\rm{eff}}$ with redshift ($Q = \delta H \rho_{\rm{de}}$)}
\end{minipage}\hspace{0.6pc}%
\begin{minipage}{17.6pc}
\includegraphics[width=17.6pc]{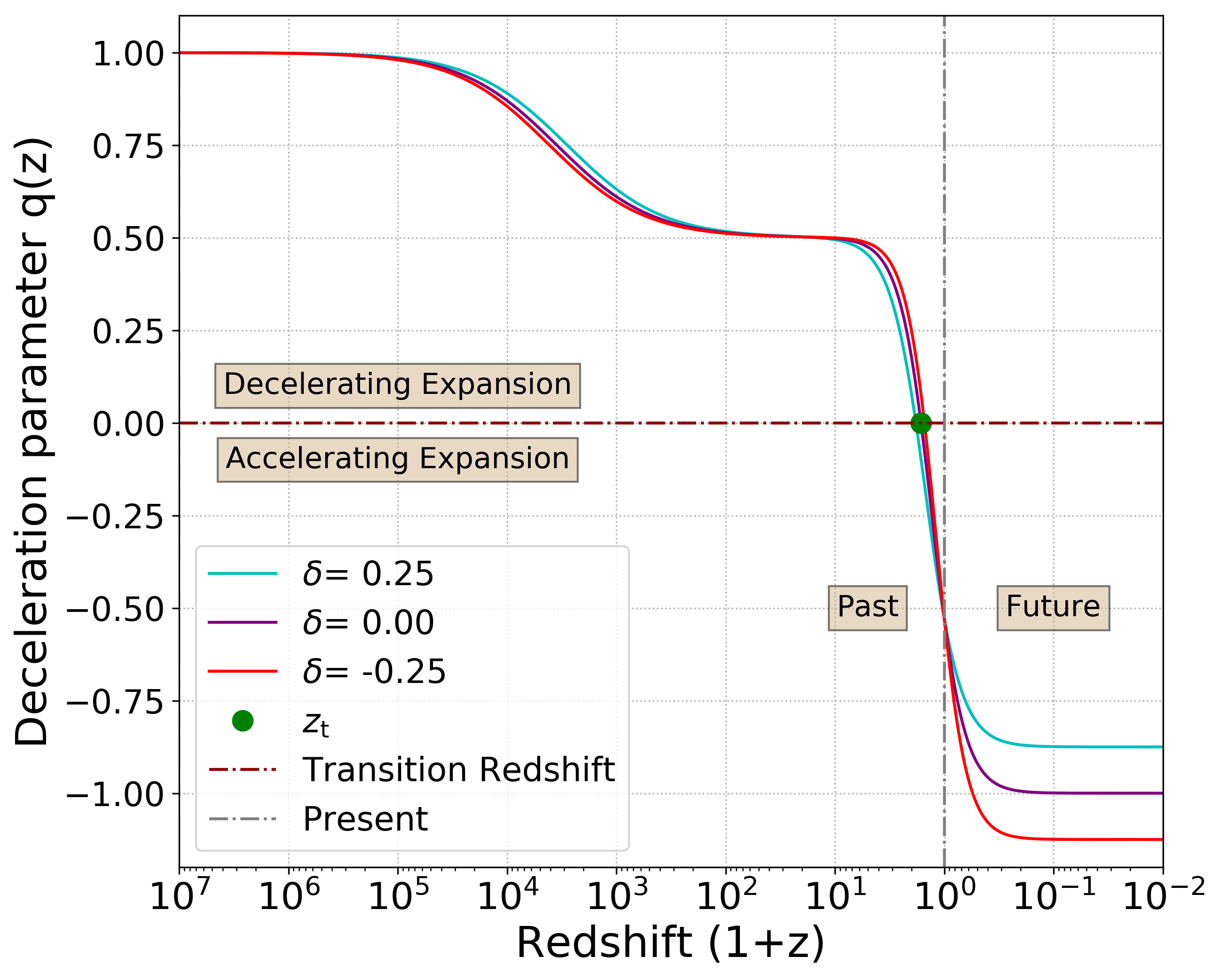} \vspace{-1.8pc}
\caption{\label{q_evo_Q2} Evolution of deceleration parameter $q$ with redshift ($1+z$) ($Q = \delta H \rho_{\rm{de}}$)}
\end{minipage}
\end{figure}
\end{center}

For this model, we have seen that the density parameters mostly deviate from the $\Lambda$CDM model during dark energy domination. We, therefore, expect this coupling function to change the behaviour of the deceleration parameter $q$ and the effective equation of state for the fluid $\omega^{\rm{eff}}$ most dramatically in the future expansion. The expressions for both $q$ (equation \ref{q}) and $\omega^{\rm{eff}}$ (equation \ref{eff}) are the same for all IDE models, with only the density parameters $\Omega_{\rm{x}}$ differing. Thus, for this model, we substitute in the density parameters for dark matter $\Omega_{\rm{dm}}$ (\ref{Q2_energy.4}), dark energy $\Omega_{\rm{de}}$ (\ref{Q2_energy.5}), radiation $\Omega_{\rm{r}}$ (\ref{Q2_energy.7}) and baryonic matter $\Omega_{\rm{bm}}$ (\ref{Q2_energy.5}) into equations (\ref{q}) and (\ref{eff}), yielding figures \ref{omega_eff_evo_Q2} and \ref{q_evo_Q2}.\vspace{-1pc}
\vspace{-2.0pc}
From figures \ref{omega_eff_evo_Q2} and \ref{q_evo_Q2}, we can see that the past behaviour for the coupled models is almost identical to that of the $\Lambda$CDM model, with initial deceleration followed by acceleration from the cosmic jerk onwards. This cosmic jerk occurs at the transition redshift $z_{\rm{t}}$, for which an analytical expression (\ref{EQ_Q2_10}) can be derived by setting $q=0$ in equation (\ref{q}), giving:
\begin{gather} \label{EQ_Q2_10}
\begin{split}
   \rightarrow       z_{\rm{t}}&=        \left( - \frac{\frac{\Omega_{\rm{(bm,0)}} +  \Omega_{\rm{(dm,0)}}}{\Omega_{\rm{(de,0)}}} +  \frac{\delta }{\delta + 3 \omega}}{1+3 \omega+  \frac{\delta }{\delta + 3 \omega}} \right)^{\left(\frac{1}{ \delta + 3 \omega } \right)} -1.
\end{split}
\end{gather}
The transition redshift for all three values of $\delta$ is calculated from (\ref{EQ_Q2_10}) and indicated by the marker in figure \ref{omega_eff_evo_Q2} and \ref{q_evo_Q2}, while the exact redshift for each can be found in tables \ref{Table_Q2_iDEDM_events} and \ref{Table_Q2_iDMDE_events}. Based on these results, we can  confirm the conclusions from table  \ref{Table_Q_events_relative}, which state that: \vspace{-0.2cm}
\begin{equation} \label{q.Q2.3}
\text{Cosmic jerk (}z_{\rm{t}} \text{)}
 \begin{cases}
 \delta>0 \text{ (iDEDM): } \quad z_{\rm{IDE}}>z_{\Lambda\rm{CDM}}  & \text{happens \textit{earlier} than }\Lambda\text{CDM}, \quad \quad \\
 \delta<0 \text{ (iDMDE): }  \quad z_{\rm{IDE}}<z_{\Lambda\rm{CDM}}  & \text{happens \textit{later} than }\Lambda\text{CDM}.
\quad \quad
 \end{cases}    
\end{equation}
From figures \ref{omega_eff_evo_Q2} and \ref{q_evo_Q2} it can also be seen that similar to the $\Lambda$CDM model, these models experience complete radiation-domination ($\Omega_{\rm{r}}, \Omega_{\rm{dm+bm}}, \Omega_{\rm{de}} ) \approx (1,0,0)\rightarrow q=1$ ; $\omega^{\rm{eff}}=1/3$, followed by complete matter-domination ($\Omega_{\rm{r}}, \Omega_{\rm{dm+bm}}, \Omega_{\rm{de}} ) \approx (0,1,0)\rightarrow q=1/2$ ; $\omega^{\rm{eff}}=0$.
As seen in figure \ref{fig_Omega_Q2}, these models do not show complete dark energy domination, but instead, the density parameters are obtained from the attractor point (\ref{DSA.3.3}), such that we have ($\Omega_{\rm{r}}, \Omega_{\rm{dm+bm}}, \Omega_{\rm{de}} ) \approx \left(0,- \frac{\delta}{ 3 \omega}, 1+\frac{\delta}{ 3 \omega} \right)$.  The deceleration parameter (\ref{q}) during dark energy domination then becomes $q_+=\frac{1}{2} \left(1 +3 \omega^{\rm{eff}}_{\rm{de}} \right)$.
For the effective equation of state (\ref{eff}) we have:
\begin{gather}
\begin{split} \label{q.Q2.5}
\omega^{\rm{eff}}_+ =\frac{\frac{1}{3}\Omega_{\rm{r}} + \omega\Omega_{\rm{de}}}{\Omega_{\rm{r}}+\Omega_{\rm{bm}}+\Omega_{\rm{dm}}+\Omega_{\rm{de}}} = \frac{\frac{1}{3}(0) + \omega\left(1+ \frac{\delta}{ 3 \omega} \right)}{(0)+(- \frac{\delta}{ 3 \omega})+\left(1+ \frac{\delta}{ 3 \omega} \right)} =  \frac{\omega+ \frac{\delta}{ 3}}{1}= \omega+ \frac{\delta}{ 3}= \omega^{\rm{eff}}_{\rm{de}},
\end{split}
\end{gather}
where (\ref{q.Q2.5}) reduces back to the $\Lambda$CDM case when either $\delta=0$ or $\omega^{\rm{eff}}_{\rm{de}}=\omega_{\rm{de}}$. We can now calculate $q_+$ and $\omega^{\rm{eff}}_+$ for the parameters used in figures \ref{omega_eff_evo_Q2} and \ref{q_evo_Q2}. Thus, for  $\delta=0.25$ (iDEDM) we have $q_+=\frac{1}{2}\left(1+3\left[-1+\frac{0.25}{3(-1)}\right]\right)=-0.875$ and $\omega^{\rm{eff}}_+=\left(-1+\frac{0.25}{3(-1)}\right)= 0.916$, while for $\delta=-0.25$ (iDMDE) we have $q_+=\frac{1}{2}\left(1+3\left[-1+\frac{0.25}{3(-1)}\right]\right)=-1.125$ and $\omega^{\rm{eff}}_+=\left(-\frac{-0.25}{3}\right)= 1.083$ for dark energy-domination. These results can be seen to exactly match the values that $q$ and $\omega^{\rm{eff}}$ converge to in figures \ref{omega_eff_evo_Q2} and \ref{q_evo_Q2} during dark energy-domination. \\
\subsubsection{Hubble parameter and age of the universe } \label{HubbleQ2}
\begin{center}
\begin{figure}[h]
\begin{minipage}{18pc}
\includegraphics[width=18pc]{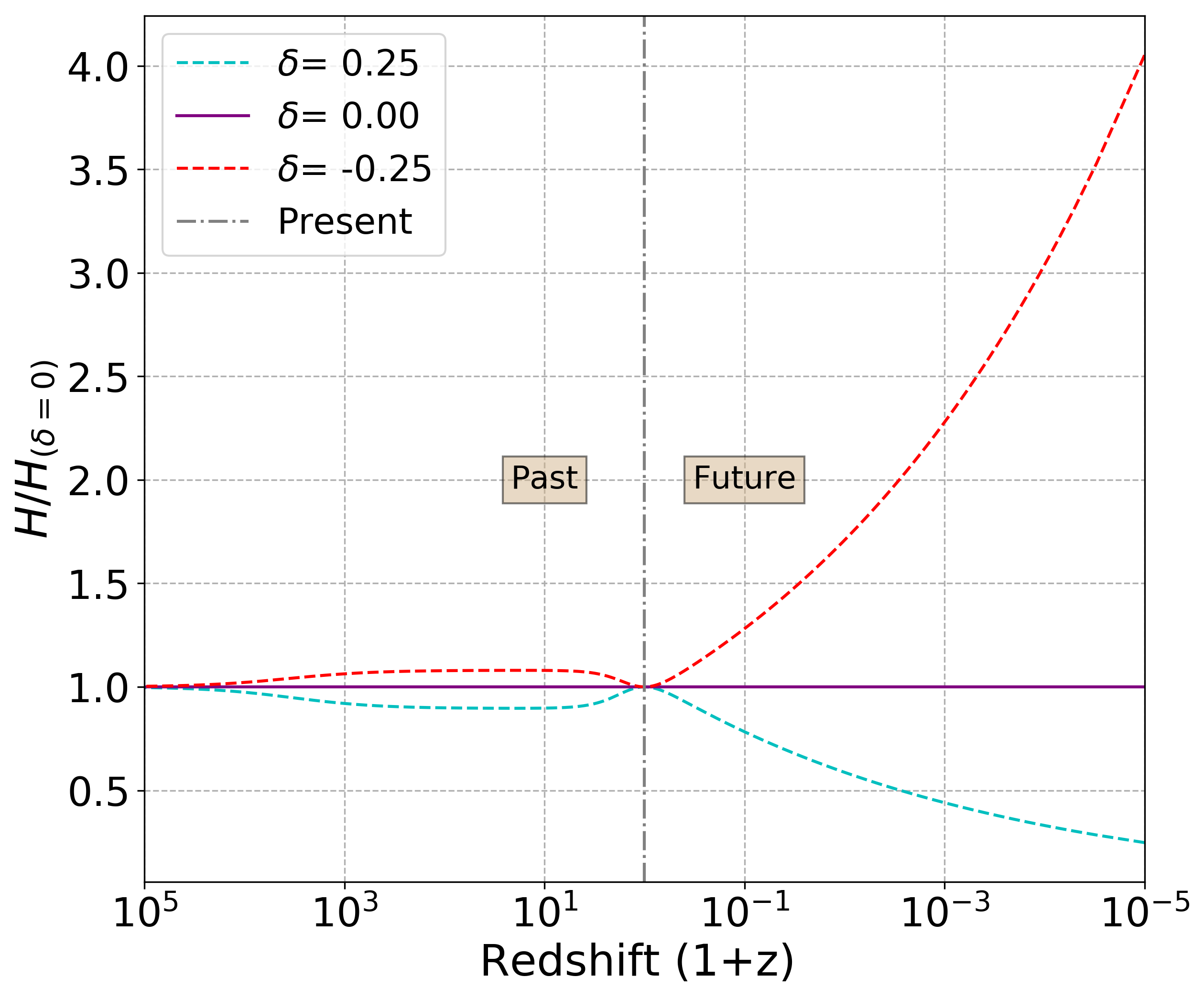}
\caption{\label{H_rel_Q2} Relative Hubble parameter ($H/H_{\delta=0}$) vs redshift ($Q= \delta H \rho_{\rm{de}}$)}
\end{minipage}\hspace{0.3pc}%
\begin{minipage}{18pc}
\includegraphics[width=18pc]{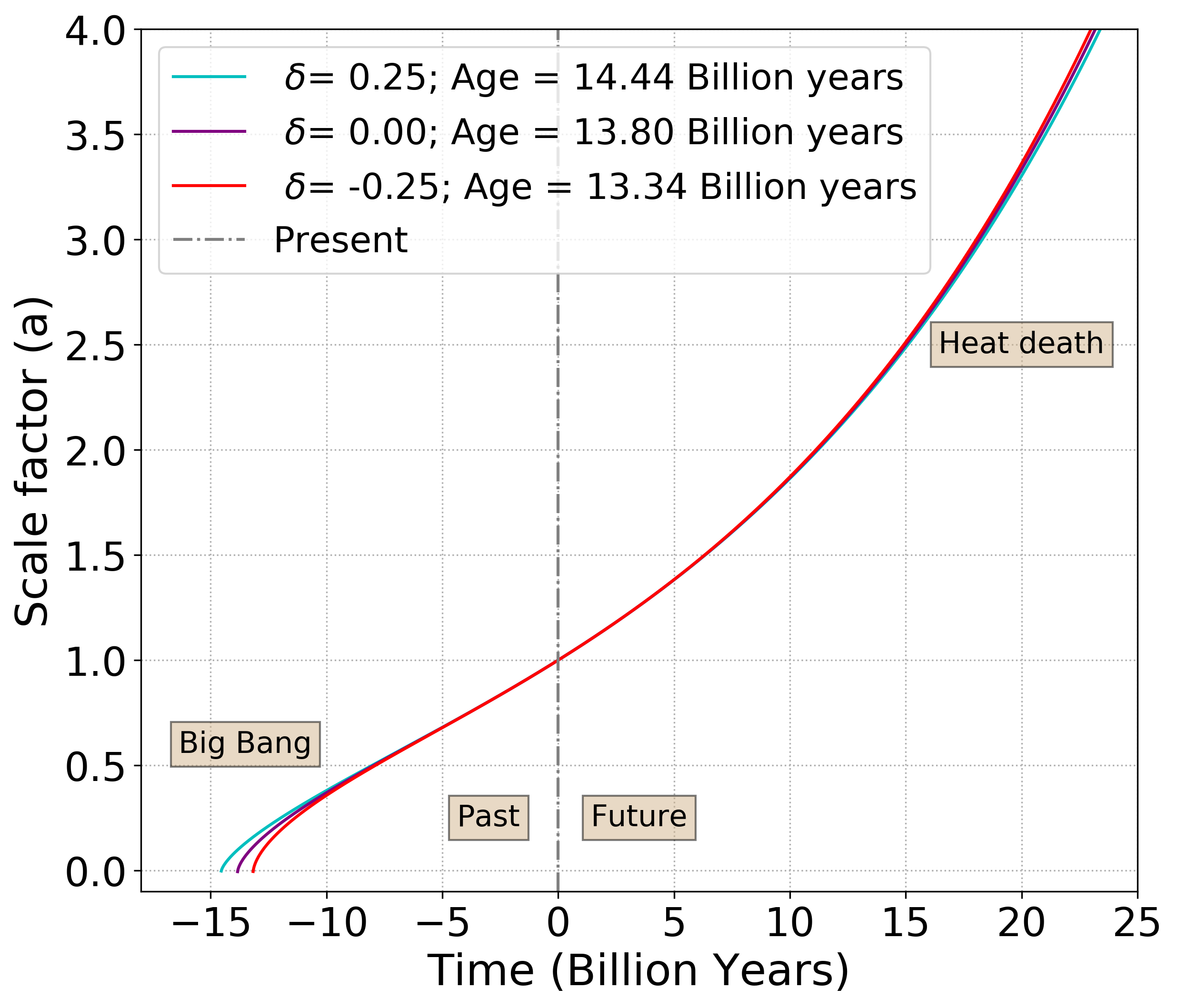}
\caption{\label{scalefactor_Q2} Evolution of scale factor with time ($Q= \delta H \rho_{\rm{de}}$)}
\end{minipage}
\end{figure}
\end{center}

The interaction $Q$ will affect the age of the universe, which can be seen from the evolution of the Friedmann equation from (\ref{H}) with the energy densities $ \rho_{\rm{r}}=\rho_{\rm{(r,0)}}a^{-4}$, $\rho_{\rm{bm}}=\rho_{\rm{(bm,0)}}a^{-3}$, $ \rho_{\rm{dm}}$ from (\ref{Q2_energy.dm}), $\rho_{\rm{de}}$ from (\ref{Q2_energy.de}) and $k=0$. Since both the deceleration parameter and the total effective equation of state deviate mostly during dark energy domination, we expect the expansion rate to change mostly for future expansion. This may be seen by plotting the Hubble parameter (\ref{H}), relative to the non-interacting case ($H/H_{\delta=0}$), against redshift. The evolution of the scale factor against time is also plotted, and the universe's age is calculated by numerically integrating (\ref{H}). This yields figures \ref {H_rel_Q2} and \ref{scalefactor_Q2}.
\vspace{-2.3pc}
In figure \ref{H_rel_Q2}, $H/H_{\delta=0}<1$ for $\delta>0$ (iDEDM) throughout most of the expansion history, indicating a slower expansion rate.
This is again due to the overall suppression of dark matter seen in figure \ref{fig_Omega_Q2}, which causes a lower value for $q$ and $\omega^{\rm{eff}}$ and therefore a slower expansion rate. Due to $\delta>0$ (iDEDM) having a slower expansion rate, more time is needed for the universe to evolve from a singularity ($a=0$) to its current size ($a=1$), causing an older age for the universe as seen in figure \ref{scalefactor_Q2}. The opposite of this holds for $\delta<0$ (iDEDM). We therefore confirm the following result from table \ref{Table_Q_events_relative}:  \vspace{-0.0cm}
\begin{equation} \label{Age.Q2.3}
\text{Age of universe (}t_{\rm{0}} \text{)}
  \begin{cases}
  \delta>0 \text{ (iDEDM): } \quad t_{\rm{(0,IDE)}}>t_{\rm{(0,}\Lambda\rm{CDM)}} & \text{\textit{Older} universe than }\Lambda\text{CDM}, \quad \quad \\
  \delta<0 \text{ (iDMDE): }  \quad t_{\rm{(0,IDE)}}<t_{\rm{(0,}\Lambda\rm{CDM)}}   & \text{\textit{Younger} universe than }\Lambda\text{CDM}.
\quad \quad
  \end{cases}     
\end{equation}
To get a sense of the magnitude of the changes in these important events, we may quantitatively describe the events in figures \ref{fig_rho_Q2}, \ref{fig_Omega_Q2}, \ref{omega_eff_evo_Q2}, \ref{q_evo_Q2}, \ref{H_rel_Q2} and \ref{scalefactor_Q2} with the cosmological parameters of the $\Lambda$CDM model from \cite{Planck2018}, with the additional parameter $\delta=0$ ($\Lambda$CDM), $\delta=0.25$ (iDEDM) and $\delta=-0.25$ (iDMDE), in tables \ref{Table_LCDM_events3}, \ref{Table_Q2_iDEDM_events} and \ref{Table_Q2_iDMDE_events} respectively.   
\begin{table}[h]
\caption{\label{Table_LCDM_events3} Important events in interacting dark energy model $\delta=0.00$ ($\Lambda$CDM) - $Q = \delta H \rho_{\rm{de}}$}  
\begin{center}
\begin{tabular}{|c|c|c|c|c|c|}
\hline
Event & Redshift  $z$ & Time (Gyr) & $\rho_{\text{r}}$  & $\rho_{\text{dm+bm}}$  & $\rho_{\Lambda}$ (J/m$^{3}$)\\
\hline
\hline
Big bang singularity & $\infty$ & 13.80  & $\infty$  & $\infty$ & $\infty$ \\
\hline
Radiation-matter equality  & 3499 & 13.80 & 10.9  & 10.9 & 5.5e-10 \\
\hline
Cosmic jerk & 0.63 & 6.12  & 5.2e-13  & 1.2e-9 & 5.5e-10 \\
\hline
Matter-dark energy equality & 0.30 & 3.50 & 2.1e-13  & 1.1e-9 & 5.5e-10 \\
\hline
\end{tabular}
\end{center}
\end{table} \vspace{-1.1cm}
\begin{table}[h]
\caption{\label{Table_Q2_iDEDM_events} Important events in interacting dark energy model $\delta=0.25$ (iDEDM) - $Q = \delta H \rho_{\rm{de}}$ }  
\begin{center}
\begin{tabular}{|c|c|c|c|c|c|}
\hline
Event & Redshift  $z$ & Time (Gyr) & $\rho_{\text{r}}$  & $\rho_{\text{dm+bm}}$  &  $\rho_{\rm{de}}$ (J/m$^{3}$)\\
\hline
\hline
Big bang singularity & $\infty$ & 14.44  & $\infty$  & $\infty$ & $\infty$ \\
\hline
Radiation-matter equality  & 2807 & 14.44 & 4.5  & 4.5 & 4.0e-9 \\
\hline
Cosmic jerk & 0.82 & 7.23  & 8.0e-13  & 1.3e-9 & 6.4e-10 \\
\hline
Matter-dark energy equality & 0.39 & 4.35  & 2.7e-13  & 6.0e-10 & 6.0e-10 \\
\hline
\end{tabular}
\end{center}
\end{table} \vspace{-1.1cm}
\begin{table}[h]
\caption{\label{Table_Q2_iDMDE_events} Important events in interacting dark energy model $\delta=-0.25$ (iDMDE) - $Q = \delta H \rho_{\rm{de}}$ }  
\begin{center}
\begin{tabular}{|c|c|c|c|c|c|}
\hline
Event & Redshift  $z$ & Time (Gyr) & $\rho_{\text{r}}$  & $\rho_{\text{m}}$  & $\rho_{\rm{de}}$ (J/m$^{3}$)\\
\hline
\hline
Big bang singularity & $\infty$ & 13.34  & $\infty$  & $\infty$ & $\infty$ \\
\hline
Radiation-matter equality  & 4084 & 13.34  & 20.3  & 20.3 & 6.6e-11 \\
\hline
Cosmic jerk & 0.52 & 5.29 & 7.8e-13  & 1.3e-9 & 6.4e-10 \\
\hline
Matter-dark energy equality & 0.24 & 2.93 & 1.7e-13 & 5.3e-10 & 5.3e-10\\
\hline
\end{tabular}\vspace{-0.4cm}
\end{center}
\end{table} \\ 

\subsubsection{Doom factor and big rip } \label{Q2rip}
\begin{center}
\begin{figure}[h]
\begin{minipage}{18.0pc}
\includegraphics[width=18.0pc]{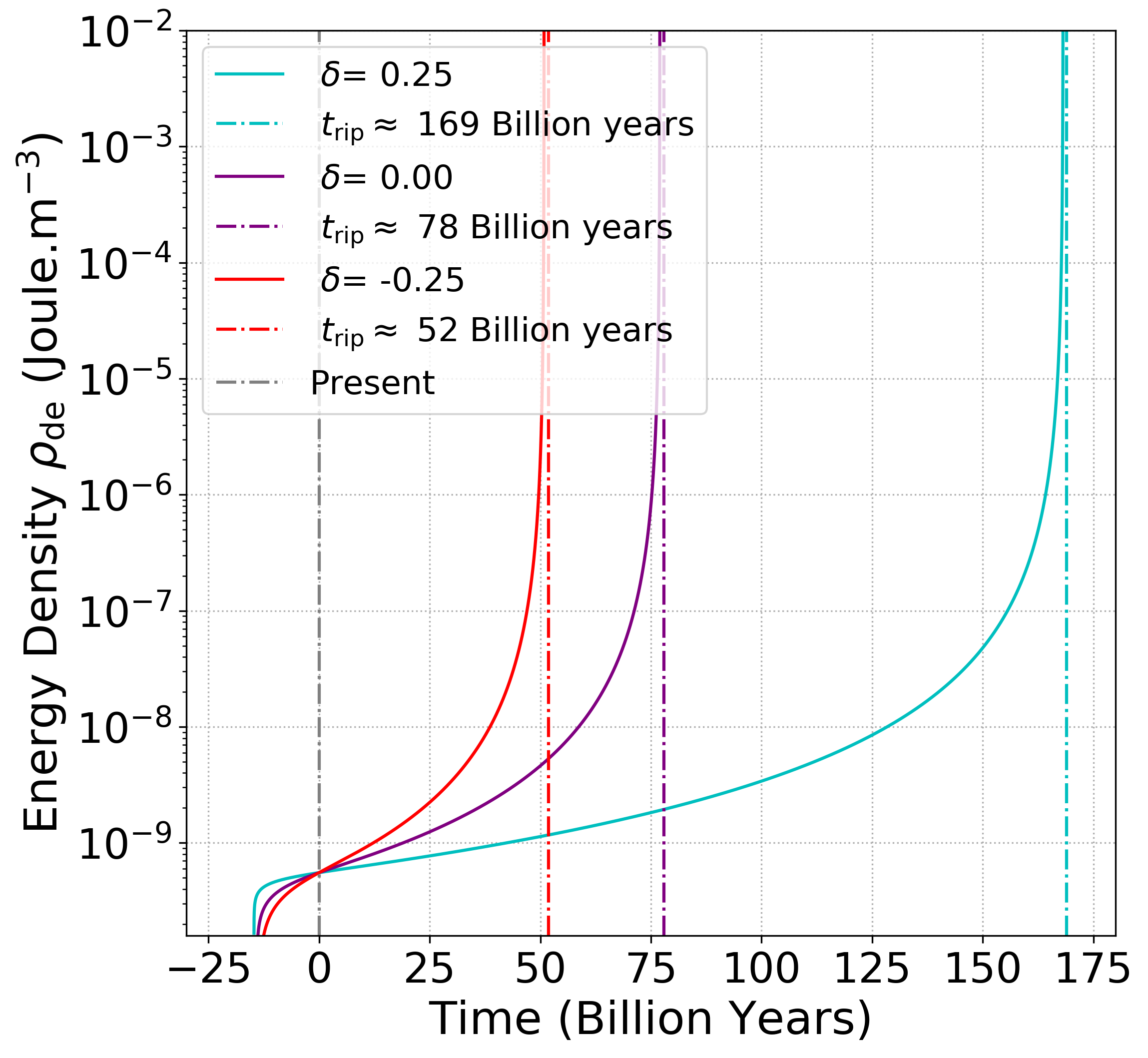}
\end{minipage}
\begin{minipage}{18.0pc} 
\includegraphics[width=18.0pc]{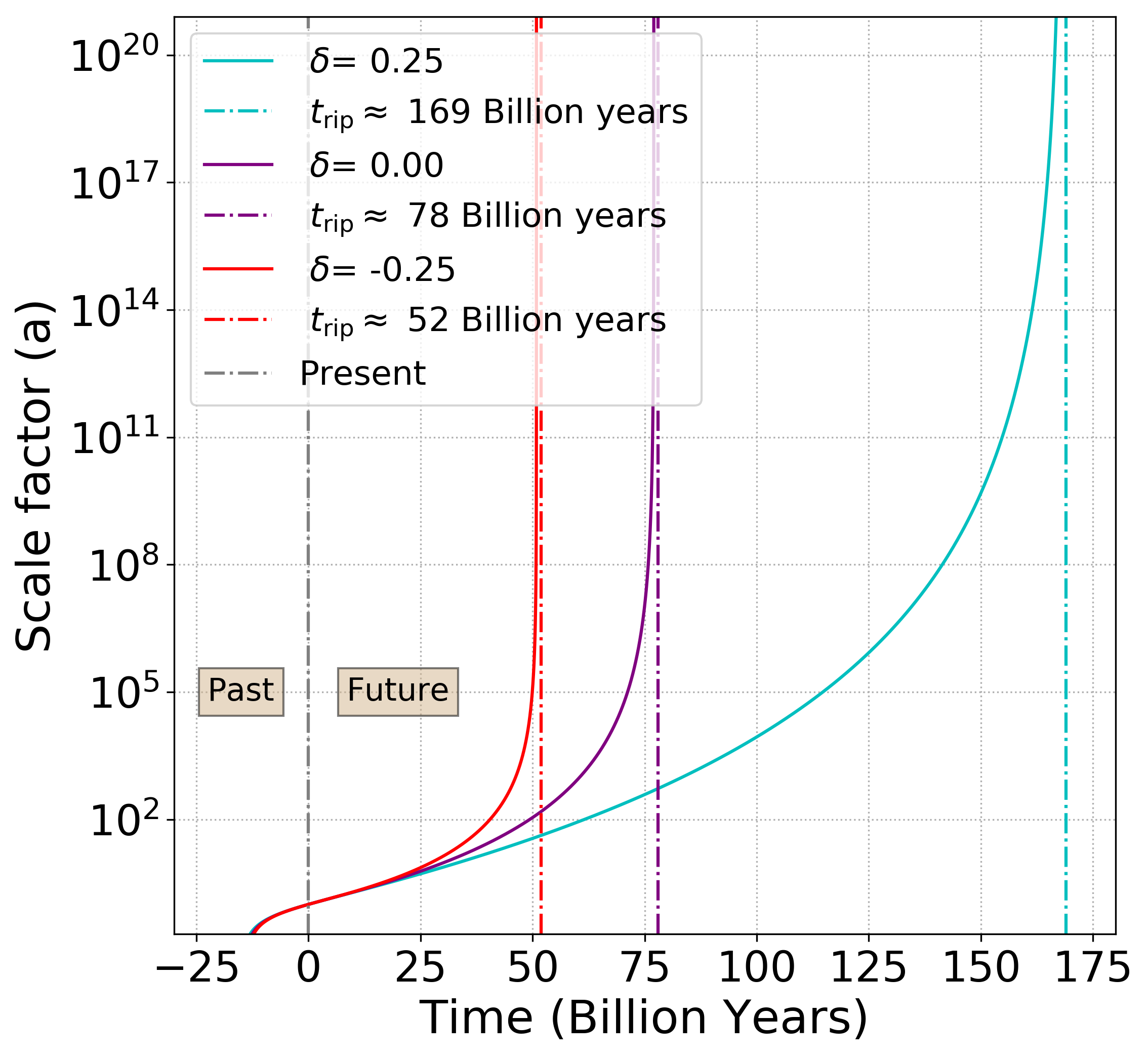}
\end{minipage}
\caption{\label{a_evo_rip_Q2} Evolution of energy density, scale factor and the big rip for phantom ($\omega=-1.15$) IDE models - ($Q = \delta H \rho_{\rm{de}}$)}
\end{figure}  
\end{center}

As previously discussed, an equation of state $\omega=-1$ causes gravitational instabilities \citep{Lucca2020, Lucca2021}. The stability of this model will once again be dependent on the doom factor \textbf{d} (\ref{d.1}). This condition $\textbf{d}<0$ guarantees an a priori stable universe as discussed in section \ref{doom}. Thus, for $Q= \delta H \rho_{\rm{de}}$ we have the doom factor (\ref{d.1}) \citep{Gavela2009}: 
\begin{align}  \label{d.2_Q2}
\textbf{d} &= \frac{Q}{3 H \rho_{\rm{de}} (1+\omega)} = \frac{\delta H \rho_{\rm{de}}}{3 H \rho_{\rm{de}} (1+\omega)} = \frac{\delta }{3  (1+\omega)},
\end{align}
where we require $\textbf{d}<0$ to ensure the stability of the universe. We can see from (\ref{d.2_Q1}), that this only occurs if $\delta$ and $(1+\omega)$ have opposite signs \citep{Gavela2009,ValentinoH02020, ValentinoCU2021, ValentinoCT2020, Lucca2020, Lucca2021}:
\begin{equation} \label{d.3_Q2}
\textbf{d}<0
  \begin{cases}
\delta <0 \quad; \quad\omega>-1\quad \text{ (Quintessence regime)}  \\
\delta >0 \quad ; \quad \omega<-1 \quad \text{ (Phantom regime)}  \\
  \end{cases} \rightarrow \text{No instabilities expected }  \quad \quad \quad \quad \quad \quad
\end{equation} 
\begin{equation} \nonumber 
\textbf{d}>0
  \begin{cases}
\delta >0 \quad; \quad\omega>-1\quad \text{ (Quintessence regime)}  \\
\delta <0 \quad ; \quad \omega<-1 \quad \text{  (Phantom regime)}  \\
  \end{cases} \rightarrow \text{Instabilities can develop if \textbf{d}$>1$}.     \quad \quad \quad
\end{equation}
Besides being stable, these models must have positive energy densities throughout the entire past and future expansion to be viable. We, therefore, need to consider the positive energy condition $ 0 <   \delta  < - 3 \omega/( 1 + \frac{1}{r_0})$ in (\ref{Q2_PEC.5}) and table \ref{Tab_PEC_Q2}. Here it was shown that we will always have $\rho_{\rm{dm}}<0$ in the future if $\delta<0$ (iDMDE), which is unphysical. The results from (\ref{d.3_Q2}) and (\ref{Q2_PEC.5}) are summarized together in table \ref{Tab_PECStabQ2} to determine the viability of the model.
\begin{table}[h]
\centering
\begin{tabular}{|c|c|c|c|c|c|c|c|c|}
\hline
$\delta$  & Energy flow &  $\omega$ & Dark energy & $\textbf{d}$ & a priori stable & $\rho_{\rm{dm}}>0$ &$\rho_{\rm{de}}>0$ & Viable \\
\hline  \hline
\rule{0pt}{12pt} + & DE $\rightarrow$ DM & $<-1$  & Phantom & -  & $\surd$ &  $\surd$ & $\surd$ & $\surd$ \\
\hline
\rule{0pt}{12pt} +  & DE $\rightarrow$ DM & $>-1$  & Quintessence & +  & X &  $\surd$ & $\surd$ & X \\
\hline
\rule{0pt}{12pt} - & DM $\rightarrow$ DE & $<-1$ & Phantom & +   & X &  X & $\surd$ & X \\
\hline
\rule{0pt}{12pt}      - & DM $\rightarrow$ DE & $>-1$ & Quintessence &-  & $\surd$ & X & $\surd$ & X \\
\hline
\end{tabular} 
\caption{Stability and positive energy criteria ($Q = \delta H \rho_{\rm{de}}$)}
 \label{Tab_PECStabQ2}
\end{table}\\
From table \ref{Tab_PECStabQ2}, we see that the only scenario that is free from both negative energy densities and instabilities is phantom dark energy $\omega<-1$ in the $\delta>0$ (iDEDM) regime. These models will thus violate many of the energy conditions of general relativity; and suffer from the consequences of doing so \citep{Carroll2003}. Since (\ref{eos.4Q2}) and (\ref{q.Q2.5}) shows that $\omega^{\rm{eff}}_+=\omega^{\rm{eff}}_{\rm{dm}}=\omega^{\rm{eff}}_{\rm{de}} =\omega+\frac{\delta}{ 3}$  in the future, the value of $\delta$ will determine if the universe model will experience a late time big rip singularity as noted by \citep{Pan2020}. For a big rip to occur, we need $\rho_{\rm{de}}\rightarrow \infty$ in a finite time. This will only occur for this model if $\rho_{\rm{de}}$ (\ref{Q2_energy.de}) increases with scale factor as the universe expands, which only happens if the effective equation of state $\omega^{\rm{eff}}_{\rm{de}} = \omega+\frac{\delta}{3}<-1$:
\begin{align}\label{BR.6Q2}
\rho_{\rm{de}} = \rho_{\rm{(de,0)}}a^{-3(1+\omega+\frac{\delta}{3})}, \quad  \quad -3\left(1+\omega+\frac{\delta}{3} \right) >0 \quad \text{ if } \quad \omega^{\rm{eff}}_{\rm{de}} = \omega+\frac{\delta}{3}<-1.
\end{align}
If condition (\ref{BR.6Q2}) is obeyed, the equivalent equation for the time of the rip $t_{\rm{rip}}$ in uncoupled phantom dark energy models \citep{Caldwell2003, HobsonTextbook} can be derived for this IDE model (see Appendix \ref{bigripApp}) as:
\begin{gather}\label{BR.5_Q2}
\begin{split}
t_{rip}&\approx -\frac{2}{3 H_0 (1+ \omega+\frac{\delta}{3})\sqrt{\left(1  -  \frac{\delta }{\delta + 3 \omega }\right)\left(1- \Omega_{\rm{(dm+bm,0)}} \right)}},  \\
\end{split}
\end{gather}
which reduces back to the uncoupled case if $\delta=0$, found in \citep{Caldwell2003, HobsonTextbook}. The predicted time of the big rip (\ref{BR.5_Q2}) is plotted alongside the evolution of the scale factor (using the Friedmann equation (\ref{H}) with a phantom dark energy equation of state $\omega=-1.15$ for illustrative purposes),  in figure \ref{a_evo_rip_Q2}. \\
In figure \ref{a_evo_rip_Q2}, we can see that the predicted time of the big rip singularity $t_{\rm{rip}}$ (\ref{BR.5_Q2}) coincides with the time at which both $a\rightarrow \infty$ and $\rho_{\rm{de}}\rightarrow \infty$. The time $t_{\rm{rip}}$ is significantly affected by the coupling, such that: 
\begin{equation} \label{q.Q2.3.b}
\text{Big rip }
  \begin{cases}
  \delta>0 \text{ (iDEDM): } \quad t_{\rm{(rip,IDE)}}>t_{\rm{(rip,}\delta=0 \rm{)}}  & \text{\textit{Later} big rip than }\delta=0,  \\
  \delta<0 \text{ (iDMDE): }  \quad t_{\rm{(rip,IDE)}}<t_{\rm{(rip,}\delta=0\rm{)}} & \text{\textit{Earlier} big rip than }\delta=0.
   \end{cases}     
\end{equation}
These models can still be viable ($\omega<-1$ in the $\delta>0$ (iDEDM) regime) and avoid a big rip, as long as condition (\ref{BR.6Q2}) is not met, such that $\omega^{\rm{eff}}_{\rm{de}} = \omega+\frac{\delta}{3}>-1$, while quintessence models with $\omega>-1$ may also have a big rip if $\omega^{\rm{eff}}_{\rm{de}}<-1$. 
Thus, if we want to avoid a big rip at $t_{\rm{rip}}$ (\ref{BR.5_Q2}), we require the condition $\omega^{\rm{eff}}_{\rm{de}} = \omega+\frac{\delta}{3}>-1$ (\ref{BR.6Q2}), which may be rewritten as $\delta>-3(\omega + 1)$. This result may be combined with the positive energy condition $ 0 <   \delta  < - 3 \omega/( 1 + \frac{1}{r_0})$ (\ref{Q2_PEC.5}) to get the following condition:
\begin{gather} \label{AllCON}
\begin{split}
    3(\omega + 1) <   \delta  < -  \frac{3 \omega}{\left( 1 + \frac{1}{r_0} \right) } \quad \text{ with } \quad \omega_{\rm{de}} <-1. \\
\end{split}  	
\end{gather}
Condition (\ref{AllCON}) describes the iDEDM regime with phantom dark energy. It ensures that all energies are positive throughout the future and past universe evolution and that early-time gravitational instabilities and late-time future big rip singularities are avoided.

\subsection{A brief summary of results for interaction model $Q = \delta H \rho_{\rm{dm}}$} \label{Q1Sum}
We will now consider an interaction model where the interaction is proportional to the dark matter density $Q \propto \rho_{\rm{dm}}$. This model is less popular in the literature than the previous model where $Q = \delta H \rho_{\rm{de}}$, which may be due to the iDMDE regime in this model having negative dark energy densities $\rho_{\rm{de}}<0$ in the past, as was pointed out in \citep{Gavela2009}). The two models may be analysed in the same manner due to the similarity of the two interaction functions. For the sake of brevity and to avoid needless repetition, we will only briefly summarise the equivalent main results of this model. A full analysis of this model with all equivalent figures and calculations shown may be found in the dissertation on which this paper is based \cite{Marcel2022D}.   \\
To obtain analytical solutions for how the dark matter $\rho_{\rm{dm}}$ and dark energy $\rho_{\rm{de}}$ densities evolve, we need to solve the conservation equations (\ref{IDE1.3}) with $Q=\delta H \rho_{\rm{dm}}$, which yields expressions for $\rho_{\rm{dm}}$ and $\rho_{\rm{de}}$:
\begin{align}
\rho_{\rm{dm}} &= \rho_{\rm{(dm,0)}} a^{\left( \delta -3 \right)},   \label{Q1_energy.dm}  \\
    \rho_{\rm{de}}  &=  \left[ \rho_{\rm{(de,0)}} + \rho_{\rm{(dm,0)}}  \frac{\delta }{\delta + 3 \omega} \left(1 - a^{\delta + 3 \omega}  \right)  \right] a^{-3    (1 + \omega)}. \label{Q1_energy.de}
\end{align}
The effective equation of state for dark matter $\omega^{\rm{eff}}_{\rm{dm}}$ and dark energy $\omega^{\rm{eff}}_{\rm{de}}$ for this model can be obtained by substituting the coupling equation $Q= \delta H \rho_{\rm{dm}}$ into (\ref{3}), yielding: 
\begin{gather} \label{eos.1Q1}
\begin{split}
\omega^{\rm{eff}}_{\rm{dm}} =  - \frac{\delta }{3 }, \quad \quad \quad \quad  \omega^{\rm{eff}}_{\rm{de}} & =  \omega_{\rm{de}} + \frac{\delta}{3 }r.
\end{split}
\end{gather}
The solutions (\ref{Q1_energy.dm}), (\ref{Q1_energy.de}) (\ref{eos.1Q1}) match with the results found in \citep{Gavela2009, Bolotin2015, Vliviita2009}. It can be seen be seen that $\omega^{\rm{eff}}_{\rm{dm}}$ is constant throughout cosmic evolution, while $\omega^{\rm{eff}}_{\rm{de}}$ is dynamic with a dependence on the coincidence problem ratio $r=\rho_{\rm{dm}}/\rho_{\rm{de}}$, which may be given in terms of redshift $z$ by the following equation:
\begin{gather} \label{r.2_Q1}
\begin{split}
r(z)  &= \frac{1 }{ \left( \frac{1}{r_0}  +   \frac{\delta }{\delta + 3 \omega} \right) (1+z)^{(\delta + 3 \omega)}  - \frac{\delta }{\delta + 3 \omega}   }, \\ 
\end{split}
\end{gather}
which matches \citep{Wang2016}. Equations (\ref{Q1_energy.dm}), (\ref{Q1_energy.de}), (\ref{eos.1Q1}), and (\ref{r.2_Q1}) can be seen to reduce back to the $\Lambda$CDM model when $\delta=0$ and $\omega=-1$. %
Similarly to what was done in section \ref{PECQ2}, we may obtain the following general condition to ensure that the energy densities will always remain positive for the coupling model $Q = \delta H \rho_{\rm{dm}}$. This condition is:
\begin{gather} \label{Q1_PEC.5}
\begin{split}
0 <   \delta  < -  \frac{3 \omega}{\left( 1 + r_0 \right) }.  \\
\end{split}  	
\end{gather}
Similar to table \ref{Tab_PEC_Q2}, the energy densities for all these conditions may be encapsulated in table \ref{Tab_PEC_Q1} below.
\begin{table}[h]
\centering
\begin{tabular}{|c|c|c|c|c|c|}
\hline
Conditions & $\rho_{\rm{dm}}$ (Past) & $\rho_{\rm{dm}}$ (Future) &  $\rho_{\rm{de}}$ (Past) & $\rho_{\rm{de}}$ (Future) & Physical\\
\hline  \hline
\rule{0pt}{13pt}     $ 0 <   \delta  < -  \frac{3 \omega}{\left( 1 + r_0 \right) }$& +& +  & + & + & $\surd$ \\
\hline
\rule{0pt}{13pt} $\delta > 0$ ; $\delta > -  \frac{3 \omega}{\left( 1 + r_0 \right) }   $& +& + & + & $-$ & X \\
\hline
\rule{0pt}{13pt} $\delta < 0$ & + & + & $-$ & + & X \\
\hline
\end{tabular}
\caption{ Conditions for positive energy densities throughout cosmic evolution $\left( Q_1 = \delta H \rho_{\rm{dm}} \right)$}
  \label{Tab_PEC_Q1}
\end{table} \\
From table \ref{Tab_PEC_Q1}, it may be seen in that in the iDMDE regime ($\delta<0$), the dark energy density always becomes negative $(\rho_{\rm{de}}<0)$ in the past, as was pointed out in \citep{Gavela2009}. This zero energy crossing $(\rho_{\rm{de}}=0)$ happens at exactly redshift $ z_{\rm{(de=0)}}$:
\begin{gather} \label{Q1_PEC.2}
\begin{split}
z_{\rm{(de=0)}} = \left[1 + \frac{1}{r_0} \left( \frac{\delta + 3 \omega}{\delta } \right)  \right]^{-\frac{1}{\delta +3 \omega}}-1.
\end{split}  	
\end{gather}
This zero energy redshift $ z_{\rm{(de=0)}}$ (\ref{Q1_PEC.2}) and the exact positive energy condition \ref{Q1_PEC.5}  may be considered new results. Therefore, from table  \ref{Tab_PEC_Q1} for the coupling $Q = \delta H \rho_{\rm{dm}}$, the iDMDE regime ($\delta<0$) should be considered unphysical, while the iDEDM ($\delta<0$) regime may be physical if condition (\ref{Q1_PEC.5}) is met. \\ 
For this model, all the results from table  \ref{Table_Q_events_relative} hold, but there is an important difference concerning how this model addresses the coincidence problem. Where $Q = \delta H \rho_{\rm{de}}$ can alleviate or even solve the coincidence problem in the future (\ref{eos.summary_Q2}), this model instead effectively addresses this during past expansion, when the interaction between the dark sector is most prominent. Similarly to (\ref{eos.summary_Q2}), this result may be summarised as: 
\begin{equation} \label{eos.summary_Q1}
 \delta >0 \; (\text{iDEDM})
   \begin{cases}
  \text{Past expansion: }  \quad \quad \omega^{\rm{eff}}_{\rm{dm}} = \omega^{\rm{eff}}_{\rm{de}} \; (\zeta_{\rm{Q}}=0) & \text{\textit{solves} coincidence problem} \\
  \text{Future expansion: } \quad \omega^{\rm{eff}}_{\rm{dm}} < \omega_{\rm{dm}} \; (\zeta_{\rm{Q}}< \zeta) & \text{\textit{alleviates} coincidence problem}, \quad \quad \quad \quad
   \end{cases}      
\end{equation}\vspace{-0.4cm}
 \hspace{-0cm} \begin{equation}  \nonumber 
   \delta <0  \; (\text{iDMDE})
   \begin{cases}
  \text{Past expansion: } \quad  \quad  \omega^{\rm{eff}}_{\rm{dm}} = \omega^{\rm{eff}}_{\rm{de}} \;(\rho_{\rm{de}}<0) &  \text{negative energy densities } \\
  \text{Future expansion: }  \quad \omega^{\rm{eff}}_{\rm{dm}}>\omega_{\rm{dm}} \; (\zeta_{\rm{Q}}> \zeta) & \text{\textit{worsens} coincidence problem}. 
   \end{cases}       \quad \quad \; \;
\end{equation} 
Furthermore, as done in section \ref{Q2rip}, we need this model to be free from gravitational instabilities. Thus, for $Q = \delta H \rho_{\rm{dm}}$ we have the doom factor (\ref{d.1}): 
\begin{align}  \label{d.2_Q1}
\textbf{d} &= \frac{Q}{3 H \rho_{\rm{de}} (1+\omega)} = \frac{\delta H \rho_{\rm{dm}}}{3 H \rho_{\rm{de}} (1+\omega)} =  \frac{\delta }{3 (1+\omega)} \frac{\rho_{\rm{dm}}}{\rho_{\rm{de}}},
\end{align}
where we also apply the conditions that $\rho_{\rm{dm}}>0$ ; $\rho_{\rm{de}} >0$. Since we need $\textbf{d}<0$ to ensure a stable universe, we can see from (\ref{d.2_Q1}) that this will only occur if $\delta$ and $(1+\omega)$ have opposite signs \citep{Gavela2009}. These results should be combined with the positive energy condition $0 <   \delta  < -  \frac{3 \omega}{\left( 1 + r_0 \right) }$ from (\ref{Q1_PEC.5}) and table \ref{Tab_PEC_Q1}. This implies that $\rho_{\rm{de}}<0$ if $\delta<0$ (iDMDE), which is unphysical and should be ruled out. The results from (\ref{d.2_Q1}) and (\ref{Q1_PEC.5}) are taken together in table \ref{Tab_PECStabQ1}.
\begin{table}[h]
\centering
\begin{tabular}{|c|c|c|c|c|c|c|c|c|}
\hline
$\delta$  & Energy flow &  $\omega$ & Dark energy & $\textbf{d}$ & a priori stable & $\rho_{\rm{dm}}>0$ &$\rho_{\rm{de}}>0$ & Viable \\
\hline  \hline
\rule{0pt}{12pt} + & DE $\rightarrow$ DM & $<-1$  & Phantom & -  & $\surd$ &  $\surd$ & $\surd$ & $\surd$ \\
\hline
\rule{0pt}{12pt} +  & DE $\rightarrow$ DM & $>-1$  & Quintessence & +  & X &  $\surd$ & $\surd$ & X \\
\hline
\rule{0pt}{12pt} - & DM $\rightarrow$ DE & $<-1$ & Phantom & +   & X &  $\surd$ & X & X \\
\hline
\rule{0pt}{12pt}      - & DM $\rightarrow$ DE & $>-1$ & Quintessence &-  & $\surd$ & $\surd$ & X & X \\
\hline
\end{tabular}
\caption{Stability and positive energy criteria ($Q_1 = \delta H \rho_{\rm{dm}}$)}
  \label{Tab_PECStabQ1}
\end{table}\\
From table \ref{Tab_PECStabQ1}, we see that the only scenario that is free from both negative energy densities and instabilities is phantom dark energy $\omega<-1$ in the $\delta>0$ (iDEDM) regime. This has the consequence that these models will also be plagued by the problems associated with phantom dark energy. Since $\omega^{\rm{eff}}_{\rm{de}}= \omega_{\rm{de}}$ (\ref{eos.summary_Q1}) in the future, an immediate consequence of dark energy being \textit{in the phantom regime, is that the universe model will always experience a late time big rip singularity} as noted by \citep{Pan2020}. 
An equivalent equation to (\ref{BR.5_Q2}) for the time of the big rip $t_{\rm{rip}}$ \citep{Caldwell2003, HobsonTextbook} is derived for this IDE model, similar to that done in Appendix \ref{bigripApp}, giving:
\begin{gather}\label{BR.5_Q1}
\begin{split}
t_{\rm{rip}} &\approx -\frac{2}{ 3 H_0 (1+ \omega) \sqrt{1-\Omega_{\rm{(bm,0)}}-\left(1-\frac{\delta}{\delta+3 \omega}\right) \Omega_{\rm{(dm,0)}}}},
\end{split}
\end{gather}  \\
which reduces back to the uncoupled case if $\delta=0$, found in \citep{Caldwell2003, HobsonTextbook}.

\section{Conclusions}
In this paper, we clarified the cosmological consequences of IDE models for any generic interaction $Q$ as summarised in table \ref{Table_Q_events_relative}. We also derived equation (\ref{DSA.8}) which may be used to obtain phase portraits for the evolution of dark matter and dark energy densities for any generic interaction $Q$ without the need to solve the conservation equation (\ref{IDE1.3}). We then considered two case studies of linear dark energy couplings, $Q = \delta H \rho_{\rm{de}}$ and $Q = \delta H \rho_{\rm{dm}}$. For these models, we derived the often neglected positive energy conditions  $ 0 <   \delta  < - 3 \omega/( 1 + \frac{1}{r_0})$ (\ref{Q2_PEC.5}) and $0 <   \delta  < -  3 \omega /\left( 1 + r_0 \right) $ (\ref{Q1_PEC.5}) respectively, from which we note the important fact that the $\delta<0$ (iDMDE) regime will always lead to $\rho_{\rm{dm}}<0$ in the future for $Q = \delta H \rho_{\rm{de}}$ and $\rho_{\rm{de}}<0$ in the past for $Q = \delta H \rho_{\rm{dm}}$. This implies that the $\delta<0$ (iDMDE) regime should not be taken seriously as a potential dark energy candidate for these models. For the more viable $\delta>0$ (iDEDM) regime,  we saw that the model $Q = \delta H \rho_{\rm{de}}$ could solve the coincidence problem in the future whilst alleviating the problem for the past (\ref{eos.summary_Q2}). Conversely, the model $Q = \delta H \rho_{\rm{dm}}$ can solve the coincidence problem in the past and alleviate the problem for the future (\ref{eos.summary_Q1}). Furthermore,  the iDEDM regime for both models predicts a \textit{later} radiation-matter equality, while both the matter-dark energy equality (\ref{eq_Q2}) and cosmic jerk will occur \textit{earlier} (\ref{q.Q2.3}). The age of these universe models will also be \textit{older} (\ref{Age.Q2.3}). The opposite holds for $\delta<0$ (iDMDE). From tables \ref{Tab_PECStabQ2} and \ref{Tab_PECStabQ1}, we see that the only viable regime for both these models, which avoid both negative energy densities and gravitational instabilities, is phantom dark energy $\omega<-1$ in the $\delta>0$ (iDEDM) regime. This has the consequence that model $Q=\delta H \rho_{\rm{dm}}$ will always end with a future big rip singularity at the derived time $t_{\rm{rip}}$ (\ref{BR.5_Q1}), while $Q = \delta H \rho_{\rm{de}}$ may avoid this fate with the right choice of cosmological parameters. The model $Q = \delta H \rho_{\rm{de}}$ will only experience a big rip future singularity at the derived time $t_{\rm{rip}}$ (\ref{BR.5_Q2}) if the condition $\omega^{\rm{eff}}_{\rm{de}} = \omega+\frac{\delta}{3}<-1$ (\ref{BR.6Q2}) is met. This big rip may be avoided if the conditions  $3(\omega + 1) <   \delta  < -  3 \omega\left( 1 + \frac{1}{r_0} \right)$, alongside $\omega_{\rm{de}} <-1$(\ref{AllCON}) is also obeyed. \\ \\
In conclusion, we are not advocating these IDE models as superior alternatives to the $\Lambda$CDM model. However, we instead want to emphasise the importance of choosing the correct parameter space (iDEDM regime, with phantom dark energy) to avoid both negative energies and instabilities (as summarised in tables \ref{Tab_PECStabQ2} and \ref{Tab_PECStabQ1}). We hope other researchers will use these theoretical constraints on the parameter space when further investigating the viability of these IDE models to address the current problems in cosmology with the latest observational data. 

\acknowledgments
MAvdW acknowledges funding through a National Astrophysical and Space Science Program (NASSP) and National Research Foundation (NRF) scholarship. AA acknowledges that this work is based on the research supported in part by the NRF and the National Institute for Theoretical and Computational Sciences of South Africa under the research theme ``New Insights into Astrophysics and Cosmology with Theoretical Models confronting Observational Data''.  

\paragraph{Masters dissertation:} 
The work presented in this article is based on the findings in the Master's dissertation of Marcel van der Westhuizen \cite{Marcel2022D}. Furthermore, an early results conference proceedings based on this work was published \cite{Marcel2022P}.
\appendix
\section{Derivation for the time of big rip in IDE model  $Q = \delta H \rho_{\rm{de}}$} \label{bigripApp}
For this model, it is important to note that in the distant future dark energy never completely dominates the other fluids (as usually indicated by $\Omega_{\rm{de}} \rightarrow 1$ in the distant future). This is because this model solves the coincidence problem for future expansion, thereby fixing the ratio of dark matter to dark energy. Radiation and baryons may become negligible in the distant future, but some terms from the dark matter energy density should still be included. The Friedmann equation (\ref{H}) for this coupled model with only dark matter (\ref{Q2_energy.dm}) and dark energy (\ref{Q2_energy.de}) is:
\begin{gather}\label{ABR2.1}
\begin{split}
\left(\frac{\dot{a}}{a} \right) &\approx H_0 \sqrt{ \left(\Omega_{\rm{(dm,0)}}    +  \Omega_{\rm{(de,0)}}  \frac{\delta }{\delta + 3 \omega } \left[ 1  -a^{- (\delta + 3 \omega) }\right]\right)a^{-3} + \Omega_{\rm{(de,0)}} a^{-3 (1 + \omega+\frac{\delta}{3})}} \\
  &= H_0 \sqrt{ \left(\Omega_{\rm{(dm,0)}}    +  \Omega_{\rm{(de,0)}}  \frac{\delta }{\delta + 3 \omega }  \right)a^{-3} + \left(1  -  \frac{\delta }{\delta + 3 \omega }\right)  \Omega_{\rm{(de,0)}}a^{-3 (1 + \omega+\frac{\delta}{3})}}.
\end{split}
\end{gather} 
In the future, as the scale factor $a$ grows large, the contribution from the first two terms in (\ref{ABR2.1}) becomes small relative to the other terms and may be neglected. Doing this, the Friedmann equation (\ref{ABR2.1}) becomes: 
\begin{align}\label{ABR2.2}
\left(\frac{\dot{a}}{a} \right) \approx H_0 \sqrt{\left(1  -  \frac{\delta }{\delta + 3 \omega }\right) \Omega_{\rm{(de,0)}} a^{-3 (1 + \omega+\frac{\delta}{3})}}.
\end{align}
The current dark energy density parameter may also be written as $\Omega_{\rm{(de,0)}}= 1 - \Omega_{\rm{(dm,0)}}- \Omega_{\rm{(bm,0)}} - \Omega_{(r,0)} \approx 1- \Omega_{\rm{(dm+bm,0)}}$. The Friedmann equation (\ref{ABR2.2}) then becomes:
\begin{gather}\label{ABR2.3}
\begin{split}
\left(\frac{\dot{a}}{a} \right) \approx H_0 \sqrt{ \left(1  -  \frac{\delta }{\delta + 3 \omega }\right)\left(1- \Omega_{\rm{(dm+bm,0)}} \right) } a^{-\frac{3}{2} (1 + \omega+\frac{\delta}{3})}.
\end{split}
\end{gather}  
This can now be integrated from the present time $t_0$ at $a_0=1$ to the time of the big rip $t_{\rm{rip}}$ at $a= \infty$:
\begin{gather}\label{ABR2.4}
\begin{split}
\frac{da}{dt} &=H_0 \sqrt{\left(1  -  \frac{\delta }{\delta + 3 \omega }\right)\left(1- \Omega_{\rm{(dm+bm,0)}} \right) } a^{-\frac{1}{2} (1 + 3\omega+\delta)}\\
\int^{t_{\rm{rip}}}_{t_0} dt &= \frac{1}{H_0}  \frac{1}{\sqrt{\left(1  -  \frac{\delta }{\delta + 3 \omega }\right)\left(1- \Omega_{\rm{(dm+bm,0)}} \right)}}  \int^{\infty}_1 a^{-\frac{1}{2} (1 + 3\omega+\delta)} da\\
t_{\rm{rip}}- t_0 &= \frac{1}{H_0} \frac{1}{\sqrt{\left(1  -  \frac{\delta }{\delta + 3 \omega }\right)\left(1- \Omega_{\rm{(dm+bm,0)}} \right)}}  \frac{2}{3 (1+ \omega+\frac{\delta}{3})} a^{\frac{3(1+\omega+\delta/3)}{2}} \Big|^\infty_{1}.
\end{split}
\end{gather}  
For the integral on the right-hand side to become zero, we need $3/2(1+ \omega +\delta/3)<0$. Phantom dark energy does not necessarily imply this. For this to hold, we need the effective state equation for dark energy to be smaller than $-1$. If $\omega^{\rm{eff}}_{\rm{de}}=\omega +\delta/3<-1$ then it follows that $3/2(1+ \omega +\delta/3)<0$ which will cause the integral $a^{\frac{3(1+\omega+\delta/3)}{2}} \approx 0$ at $a = \infty$. Thus (\ref{ABR2.4}) becomes: 
\begin{gather}\label{ABR.6}
\begin{split}
t_{\rm{rip}}- t_0 &= \frac{2}{3 H_0 (1+ \omega+\frac{\delta}{3})\sqrt{\left(1  -  \frac{\delta }{\delta + 3 \omega }\right)\left(1- \Omega_{\rm{(dm+bm,0)}} \right)}}  \left(0 - 1^{\frac{3(1+\omega+\delta/3)}{2}} \right) \\
\rightarrow t_{\rm{rip}}- t_0 &= -\frac{2}{3 H_0 (1+ \omega+\frac{\delta}{3})\sqrt{\left(1  -  \frac{\delta }{\delta + 3 \omega }\right)\left(1- \Omega_{\rm{(dm+bm,0)}} \right)}},  \\
\end{split}
\end{gather}  
which is the predicted time of the big rip for the IDE model $Q = \delta H \rho_{\rm{de}}$. This reduces back to the uncoupled case if $\delta=0$, found in \cite{HobsonTextbook, Caldwell2003}. 



\begin{thebibliography}{99}

\bibitem{Planck2018}
N. Aghanim, Y. Akrami, M. Ashdown, et al., \emph{Planck 2018 results. VI. Cosmological parameters}, \emph{Astron. Astrophys.}, {\bf 641} A6 (2020) [\href{https://doi.org/10.48550/arXiv.1807.06209}{arXiv:1807.06209} [astro-ph]]. 

\bibitem{Riess1998}
A.G. Riess, A. V. Filippenko, P. Challis, et al.,  \emph{Observational evidence from supernovae for an accelerating universe and a cosmological
constant}, \emph{Astron. J.}, {\bf 116} (1998) 1009-1038 [\href{http://dx.doi.org/10.1086/300499}{astro-ph/9805201}]. 

\bibitem{Perlmutter1999}
S. Perlmutter, G. Aldering, G. Goldhaber, et al.,  \emph{Measurements of $\omega$ and $\lambda$ from 42 highredshift supernovae}, \emph{Astrophys. J.}, {\bf 517} (1999) 565-586 [\href{http://dx.doi.org/10.1086/307221}{astro-ph/9812133}].

\bibitem{kids2021}
M. Asgari, et al., \emph{KiDS-1000 cosmology: Cosmic shear constraints and comparison between two point statistics}, \emph{Astron. Astrophys.}, {\bf 645} (2021) A104 [\href{https://doi.org/10.1051/0004-6361/202039070}{arXiv:2007.15633} [astro-ph]].

\bibitem{DES2022}
T. M. C. Abbott, et al. (DES Collaboration),  \emph{Dark Energy Survey Year 3 results: Cosmological constraints from galaxy clustering and weak lensing}, \emph{Phys. Rev. D}, {\bf 105} (2022) 023520 [\href{https://link.aps.org/doi/10.1103/PhysRevD.105.023520}{arXiv:2105.13549} [astro-ph]].

\bibitem{Weinberg1989}
S. Weinberg, \emph{The cosmological constant problem}, \emph{Rev. Mod. Phys.}, {\bf 61} (1989) 1-23 [\url{https://link.aps.org/doi/10.1103/RevModPhys.61.1}].

\bibitem{HobsonTextbook}
M. P. Hobson, G. P. Efstathiou, and A. N. Lasenby, \emph{General Relativity}, \emph{Cambridge University Press}, (2021). 

\bibitem{delCampo2009}
S. del Campo, R. Herrera, and D. Pavon, \emph{Interacting models may be key to solve the cosmic coincidence problem}, \emph{J. Cosmol. Astropart. Phys.}, {\bf 01} (2009) 020 [\href{http://dx.doi.org/10.1088/1475-7516/2009/01/020}{arXiv:0812.2210} [gr-qc]].

\bibitem{Huey2006}
G.Huey and B.D. Wandelt, \emph{Interacting Quintessence, the Coincidence Problem and Cosmic Acceleration}, \emph{Phys. Rev. D}, {\bf 74} (2006) 023519 [\href{https://link.aps.org/doi/10.1103/PhysRevD.74.023519}{astro-ph/0407196}].

\bibitem{Velten2014}
H.E.S. Velten, R.F. vom Marttens and W. Zimdahl, \emph{Aspects of the cosmological "coincidence problem"}, \emph{Eur. Phys. J. C}, {\bf 74} (2014) 3160 [\href{https://doi.org/10.1140/epjc/s10052-014-3160-4}{arXiv:1410.2509} [astro-ph]]. 

\bibitem{Wang2016} 
B. Wang, E. Abdalla, F. Atrio-Barandela, and D. Pavn, \emph{Dark matter and dark energy interactions: theoretical challenges, cosmological implications and observational signatures}, \emph{Rept. Prog. Phys.}, {\bf 79} (2016) 096901 [\href{http://dx.doi.org/10.1088/0034-4885/79/9/096901}{arXiv:1603.08299} [astro-ph]].
 
\bibitem{Bolotin2015}
Y. L. Bolotin, A. Kostenko, O. A. Lemets, and D. A. Yerokhin, \emph{Cosmological evolution with interaction between dark energy and dark matter}, \emph{Int. J. Mod. Phys. D}, {\bf 24} (2015) 1530007 [\href{http://dx.doi.org/10.1142/S0218271815300074}{arXiv:1310.0085} [astro-ph]]. 
 
\bibitem{Zlatev1999}
I. Zlatev, L. Wang, and P. J. Steinhardt, \emph{Quintessence, cosmic coincidence, and the cosmological constant}, \emph{Phys. Rev. Lett.}, {\bf 82} (1999) 896899 [\href{http://dx.doi.org/10.1103/PhysRevLett.82.896}{astro-ph/9807002}].

\bibitem{Riess2019}
A.G. Riess, S. Casertano, W. Yuan, et al., \emph{Large magellanic cloud cepheid standards provide a $1 \%$ determination of the hubble constant and stronger evidence for physics beyond $\Lambda$CDM}, \emph{Astrophys. J.}, {\bf 876} (2019) (1)85 [\href{http://dx.doi.org/10.3847/1538-4357/ab1422}{	arXiv:1903.07603} [astro-ph]].

\bibitem{Riess2021}
A.G. Riess, S. Casertano, W. Yuan, et al., \emph{Cosmic distances calibrated to $1 \%$ precision with Gaia EDR3 parallaxes and Hubble space telescope photometry of 75 milky way cepheids confirm tension with $\Lambda$CDM}, \emph{Astrophys. J.}, {\bf 908} (2021) L6 [\href{http://dx.doi.org/10.3847/2041-8213/abdbaf}{arXiv:2012.08534} [astro-ph]].

\bibitem{Dainotti2022}
M. G. Dainotti, B. De Simone, T. Schiavone, et al., \emph{On the evolution of the Hubble constant with the SNe Ia Pantheon Sample and Baryon Acoustic Oscillations: a feasibility study for GRB-cosmology in 2030
}, \emph{Galaxies}, {\bf 10} (2022) 1 [\href{https://www.mdpi.com/2075-4434/10/1/24}{arXiv:2201.09848} [astro-ph]].

\bibitem{Dainotti2021}
M. G. Dainotti, B. De Simone, T. Schiavone, et al., \emph{On the Hubble Constant Tension in the SNe Ia Pantheon Sample}, \emph{Astrophys. J.}, {\bf 912} (2021) 150 [\href{https://iopscience.iop.org/article/10.3847/1538-4357/abeb73}{arXiv:2103.02117} [astro-ph]].

\bibitem{VagnozziV22020}
S. Vagnozzi, \emph{New physics in light of the $H_0$ tension: an alternative view}, \emph{Phys. Rev. D }, {\bf 102} (2020) 023518 [\href{https://journals.aps.org/prd/abstract/10.1103/PhysRevD.102.023518}{arXiv:1907.07569} [astro-ph]].

\bibitem{ValentinoH02017}
E. Di Valentino, A. Melchiorri, O. Mena, \emph{Can interacting dark energy solve the $H_0$ tension?}, \emph{Phys. Rev. D}, {\bf 96} (2017) 043503 [\href{https://doi.org/10.1103/PhysRevD.96.043503}{arXiv:1704.08342} [astro-ph]].

\bibitem{ValentinoH02020}
E. Di Valentino, A. Melchiorri, O. Mena and S. Vagnozzi, \emph{Interacting dark energy in the early $2020$s: A promising solution to the $H_0$ and cosmic shear tensions}, \emph{Phys. Dark Universe}, {\bf 30} (2020) 100666 [\href{http://dx.doi.org/10.1016/j.dark.2020.100666}{arXiv:1908.04281} [astro-ph]]. 

\bibitem{ValentinoH02021}
E. Di Valentino, O. Mena, S. Pan, et al., \emph{In the realm of the hubble
tension $-$ a review of solutions}, \emph{Class. Quantum Gravity}, {\bf 38} (2021) 153001 [\href{https://doi.org/10.1088/1361-6382/ac086d}{arXiv:2103.01183} [astro-ph]].

\bibitem{Cyr-Racine2021}
F. Cyr-Racine, \emph{Cosmic expansion: A mini review of the hubble-lemaitre tension}, (2021) [\href{https://arxiv.org/abs/2105.0940}{arXiv:2105.09409}[astro-ph]].

\bibitem{Nunes2022}
R. C. Nunes, S. Vagnozzi, S. Kumar, E.Di Valentino, and O. Mena, \emph{New tests of dark sector interactions from the full-shape galaxy power spectrum}, \emph{Phys. Rev. D}, {\bf 105} (2022) 123506 [\href{https://link.aps.org/doi/10.1103/PhysRevD.105.123506}{arXiv:2203.08093} [astro-ph]].

\bibitem{VagnozziV22018}
W. Yang, S. Pan, E. Di Valentino, et al., \emph{Tale of stable interacting dark energy, observational signatures, and the $H_0$ tension}, \emph{J. Cosmol. Astropart. Phys.}, {\bf 09} (2018) 019 [\href{https://iopscience.iop.org/article/10.1088/1475-7516/2018/09/019}{arXiv:1805.08252} [astro-ph]].

\bibitem{LWang02022}
L. Wang, J. Zhang, D. He, J. Zhang and X. Zhang, \emph{Constraints on interacting dark energy models from time-delay cosmography with seven lensed quasars}, \emph{Mon. Notices Royal Astron. Soc.}, {\bf 514} (2022) 1433–1440 [\href{ https://doi.org/10.1093/mnras/stac1468}{arXiv:2102.09331} [astro-ph]].


\bibitem{Gariazzo2022}
S. Gariazzo, E. Di Valentino, O. Mena, and R. C. Nunes, \emph{Late-time interacting cosmologies and the Hubble constant tension}, \emph{Phys. Rev. D}, {\bf 106} (2022) 023530 [\href{https://link.aps.org/doi/10.1103/PhysRevD.106.023530}{arXiv:2111.03152} [astro-ph]].

\bibitem{Lucca2021}
M. Lucca, \emph{Dark energy-dark matter interactions as a solution to the $s_8$ tension}, \emph{Phys. Dark Universe}, {\bf 34} (2021) 100899 [\href{https://doi.org/10.1016/j.dark.2021.100899}{arXiv:2105.09249} [astro-ph].

\bibitem{NunesS82021}
R. C. Nunes and S.Vagnozzi, \emph{Arbitrating the S8 discrepancy with growth rate measurements from Redshift-Space Distortions}, \emph{Mon. Notices Royal Astron. Soc.}, {\bf 505} (2021) 5427-5437 [\href{https://doi.org/10.1093/mnras/stab1613}{arXiv:2106.01208} [astro-ph]].

\bibitem{ValentinoS82021}
E. Di Valentino, L. A. Anchordoqui, O. Akarsu, et al., \emph{Cosmology intertwined III: $f \sigma_8$ and $S_8$}, \emph{J. Cosmol. Astropart. Phys.}, {\bf 131} (2021) 102604 [\href{https://doi.org/10.1016/j.astropartphys.2021.102604}{arXiv:2008.11285} [astro-ph]].

\bibitem{Carroll2021}
S.M. Carroll, \emph{The quantum field theory on which the everyday world supervenes}, \emph{arXiv}, (2021) [\href{https://arxiv.org/abs/2101.07884}{arXiv:2101.07884} [physics.hist-ph]].

\bibitem{Bhmer2008}
C. G. Bhmer, G. Caldera-Cabral, R. Lazkoz, and R. Maartens., \emph{Dynamics of dark energy with a coupling to dark matter}, \emph{Phys. Rev. D}, {\bf 78} (2008) 023505 [\href{http://dx.doi.org/10.1103/PhysRevD.78.023505}{arXiv:0801.1565} [gr-qc]].

\bibitem{Vliviita2009}
J. Vliviita, E. Majerotto, and R. Maartens, \emph{Large-scale instability in interacting dark energy and dark matter fluids}, \emph{J. Cosmol. Astropart. Phys.}, {\bf 07} (2008) 020 [\href{http://dx.doi.org/10.1088/1475-7516/2008/07/020}{arXiv:0804.0232} [astro-ph]].

\bibitem{Gavela2009}
M.B. Gavela, D. Hernandez, L. Lopez Honorez, et al., \emph{Dark coupling}, \emph{J. Cosmol. Astropart. Phys.}, {\bf 07} (2009) 034 [\href{https://doi.org/10.1088/1475-7516/2009/07/034}{arXiv:0901.1611} [astro-ph]].

\bibitem{Gavela2010}
M.B Gavela, L. Lopez Honorez, O Mena, and S Rigolin, \emph{Dark coupling and gauge invariance}, \emph{J. Cosmol. Astropart. Phys.}, {\bf 11} (2010) 044 [\href{http://dx.doi.org/10.1088/1475-7516/2010/11/044}{arXiv:1005.0295} [astro-ph]].

\bibitem{ValentinoCU2021}
E. Di Valentino, A. Melchiorri, O. Mena, S. Pan, and W. Yang, \emph{Interacting dark energy in a closed universe}, \emph{Mon. Notices Royal Astron. Soc.: Letters}, {\bf 502}(01) (2021) L23-L28 [\href{http://dx.doi.org/10.1093/mnrasl/slaa207}{arXiv:2011.00283} [astro-ph]].

\bibitem{ValentinoCT2020}
E. Di Valentino, A. Melchiorri, O. Mena, and S. Vagnozzi, \emph{Nonminimal dark sector physics and cosmological tensions}, \emph{Phys. Rev. D}, {\bf 101}(6) (2020) 2470-0029 [\href{http://dx.doi.org/10.1103/PhysRevD.101.063502}{arXiv:1910.09853} [astro-ph]].

\bibitem{Lucca2020}
M. Lucca and D. C. Hooper, \emph{Shedding light on dark matter-dark energy interactions}, \emph{Phys. Rev. D}, {\bf 102}(12) (2020) 2470-0029 [\href{https://arxiv.org/abs/2002.06127}{arXiv:2002.06127} [astro-ph]].

\bibitem{Pan2020}
S. Pan, J. de Haro, W. Yang, and J. Amors, \emph{Understanding the phenomenology of interacting dark energy scenarios and their theoretical bounds}, \emph{Phys. Rev. D}, {\bf 101}(12) (2020) 2470-0029 [\href{http://dx.doi.org/10.1103/PhysRevD.101.123506}{arXiv:2001.09885} [gr-qc]].



\bibitem{PanField2020}
S. Pan, G.S. Sharov, and W. Yang \emph{Field theoretic interpretations of
interacting dark energy scenarios and recent observations}, \emph{Phys. Rev. D}, {\bf 101}(10) (2020) 2470-0029 [\href{http://dx.doi.org/10.1103/PhysRevD.101.103533}{arXiv:2001.03120}  [astro-ph]].

\bibitem{Carroll2003}
S. M. Carroll, M. Hoffman, and M. Trodden \emph{Can the dark energy equation-of-state parameter $\omega$ be less than $-1$?}, \emph{Phys. Rev. D}, {\bf 68} (2003) 023509 [\href{https://doi.org/10.1103/PhysRevD.68.023509}{astro-ph/0301273}].

\bibitem{Caldwell2003}
R.R. Caldwell, M. Kamionkowski, and N. N. Weinberg \emph{Phantom energy and
cosmic doomsday}, \emph{Phys. Rev. Lett.}, {\bf 91}(7) (2003) 1079-7114 [\href{http://dx.doi.org/10.1103/PhysRevLett.91.071301}{astro-ph/0302506}].

\bibitem{Marcel2022D}
M. A. van der Westhuizen, \emph{Dark interactions beyond the $\Lambda$CDM model}, Masters dissertation, North-West University (2022) [\url{https://repository.nwu.ac.za/handle/10394/39596}].

\bibitem{Marcel2022P}
M. A. van der Westhuizen, A. Abebe, \emph{Dark coupling: cosmological implications of interacting dark energy and dark matter fluids}, \emph{SA Inst. Phys.} Proceedings, ISBN: 978-0-620-97693-0, SAIP2021 (2022) 386-391 [\url{http://saip.org.za/Proceedings/Track\%20D/63.pdf}].


\end{thebibliography}
\end{document}